\documentclass[pra,twocolumn,showkeys,showpacs]{revtex4} %
\usepackage{amsmath,bm}
\usepackage{amssymb}

\renewcommand{\a}{\mathbf{a}}
\newcommand{\A}{\mathbf{A}}
\newcommand{\B}{\mathbf{B}}
\renewcommand{\b}{\mathbf{b}}
\renewcommand{\d}{\mathbf{d}}
\newcommand{\E}{\mathbf{E}}
\newcommand{\D}{\mathbf{D}}

\newcommand{\e}{\mathbf{e}}
\renewcommand{\i}{\mathbf{i}}
\newcommand{\p}{\mathbf{p}}
\renewcommand{\r}{\mathbf{r}}
\newcommand{\s}{\mathbf{s}}

\renewcommand{\v}{\mathbf{v}}
\newcommand{\z}{\mathbf{z}}
\newcommand{\x}{\mathbf{x}}
\renewcommand{\x}{\mathbf{x}}

\newcommand{\tpsi}{\widetilde\psi}

\newcommand{\tF}{\widetilde F}

\newcommand{\tN}{\widetilde N}
\newcommand{\tM}{\widetilde M}

\newcommand{\tR}{\widetilde R}
\newcommand{\tU}{\widetilde U}
\newcommand{\tV}{\widetilde V}
\newcommand{\tPsi}{\widetilde \Psi}

\newcommand{\bsig}{\bm{\sigma}}

\newcommand{\half}{{\textstyle \frac{1}{2}}}
\newcommand{\smallhalf}{{\scriptstyle\frac{1}{2}}}
\newcommand{\bdot}{\bm{\cdot}}

\begin{document}

\title{ Quantum Mechanics of the electron particle-clock  }

\author{David Hestenes}
\affiliation{Department of Physics, Arizona State University, Tempe, Arizona 85287-1504}
\email{hestenes@asu.edu}
\homepage{http://geocalc.clas.asu.edu/}

\begin{abstract}

Understanding the electron clock and the role of complex numbers in quantum mechanics
is grounded in the geometry of spacetime, and best expressed with \textit{Spacetime Algebra} (STA).
The efficiency of STA is demonstrated with coordinate-free applications to both relativistic and non-relativistic QM.
Insight into the structure of  Dirac theory is provided by 
a new comprehensive analysis of \textit{local observables} and conservation laws.
This provides the foundation for a comparative analysis of wave and particle models for
the hydrogen atom,  the workshop where Quantum Mechanics was designed, built and tested.

\end{abstract}


\maketitle

\section{Introduction}

De Broglie always insisted that Quantum Mechanics is a relativistic theory. Indeed, it began with de Broglie`s relativistic model for an electron clock.
Ironically, when Schr\"odinger introduced de Broglie`s idea  into his wave equation, he dispensed with both the clock and the relativity!

This paper aims to revitalize de Broglie's idea of an electron clock by giving it a central role in physical interpretation of the Dirac wave function.
In particular, we aim for insight into  structure of the wave function and fibrations of particle paths it determines.
This opens up new  questions about physical interpretation.

We begin with a synopsis of \textit{Spacetime Algebra} (STA), which is an essential tool in all that follows.

Section III applies STA in a review of \textit{Real Dirac theory} and provides a rigorous new analysis of its physical interpretation in terms of local observables. That facilitates the study of electron paths in subsequent sections.

Section IV discusses a new approach to Born's statistical interpretation of the Dirac wave function dubbed \textit{Born-Dirac} theory. It includes a relativistic extension of de Broglie-Bohm Pilot Wave theory to interpret the Dirac wave function as describing a fibration (or ensemble)  of possible particle paths. Spin dependence of the so-called \textit{Quantum potential} is made explicit and generalized.

Section V presents a new relativistic particle model for electron states in the hydrogen atom. This corrects errors and deficiencies of Old Quantum Theory to make it  consistent with the Dirac equation.
It invites comparison with the standard Darwin model for hydrogen discussed in Section IVD. Thereby it opens up new possibilities for experimental test and theoretical analysis.

Section VI discusses modifications of Dirac theory to fully incorporate electron zitter. This serves as an introduction to a more definitive model of the electron clock developed in a subsequent paper \cite{Hest19b}. 

The final Section argues for a realist interpretation of the Dirac wave function describing possible particle paths.

\section{Spacetime Algebra}\label{secII}

\textit{Spacetime Algebra} (STA) plays an essential role in the formulation and analysis of electron theory in this paper.
Since thorough expositions of STA are available in many places \cite{Hest03b,Doran03,Hest66}, a brief description will suffice here, mainly to establish notations and define terms.

STA is an associative algebra generated by spacetime vectors with the property that
the square of any vector is a (real) scalar. Thus for any vector $a$
we can write
\begin{equation}
a^2 = a a  = \varepsilon|a|^2\,,\label{2.1}
\end{equation}
where $\varepsilon$ is the \textit{signature} of $a$ and $|a|$ is a
positive scalar. As usual, we say that $a$ is \textit{timelike},
\textit{lightlike} or \textit{spacelike} if its signature is positive
($\varepsilon = 1$), null ($\varepsilon = 0$), or negative ($\varepsilon =-1$).

 From the \textit{geometric product} $ab$
of two vectors it is convenient to define two other
products. The \textit{inner product} $a\bdot b$ is defined by
\begin{equation}
a\bdot b = \half(ab + ba)=b\bdot a\,,\label{2.2}
\end{equation}
while the \textit{outer product } $a\wedge b$ is defined by
\begin{equation}
a\wedge b = \half(ab - ba) = - b\wedge a\,.\label{2.3}
\end{equation}
The three products are therefore related by
\begin{equation}
ab =a\bdot b + a\wedge b\,.\label{2.4}
\end{equation}
This can be regarded as a decomposition of the
product $ab$ into symmetric and skewsymmetric parts, or
alternatively, into scalar and bivector parts.

\def\g{\hbox{g}}

For physicists unfamiliar with STA, it will be helpful to note its isomorphism to Dirac algebra over the reals.
To that end, let $\{\gamma_\mu;\ 0,1,2,3\}$
be a \textit{right-handed orthonormal frame} of vectors
with $\gamma_0$  in the forward light cone.
The symbols $\gamma_\mu$ have been selected to emphasize direct correspondence with Dirac's $\gamma$-matrices.
In accordance with (\ref{2.2}), the
components $\g_{\mu\nu}$   of the metric tensor are given by
\begin{equation}
\g_{\mu\nu}=\gamma_\mu\bdot\gamma_\nu
=\half(\gamma_\mu\gamma_\nu+\gamma_\nu\gamma_\mu)\,.\label{2.5}
\end{equation}
This will be recognized as isomorphic to a famous formula
of Dirac's. Of course, the difference here is that the $\gamma_\mu$
are vectors rather than matrices.

The unit pseudoscalar $i$  for spacetime is related
to the frame $\{\gamma_\nu\}$ by the equation
\begin{equation}
i=\gamma_0\gamma_1\gamma_2\gamma_3
=\gamma_0\wedge\gamma_1\wedge\gamma_2\wedge\gamma_3\,.\label{2.6}
\end{equation}
It is readily verified from (\ref{2.6}) that $i^2=-1$,
and the geometric product\ of $i$ with any vector is
anticommutative.

By multiplication the $\gamma_\mu$ generate a
complete basis of $k$-vectors for STA, consisting
of the $2^4 = 16$ linearly independent elements
\begin{equation}
1,\quad\gamma_\mu,\quad\gamma_\mu
    \wedge\gamma_\nu,\quad\gamma_\mu i,\quad i\,.\label{2.7}
\end{equation}
Obviously, this set corresponds to 16 base matrices for the Dirac algebra, with the pseudoscalar $i$ corresponding to the Dirac matrix $\gamma_5$.

The entire spacetime algebra is obtained from
linear combinations of basis $k$-vectors in (\ref{2.7}).
A generic element $M$ of the STA, called a \textit{multivector},
can thus be written in the \textit{expanded form}
\begin{equation}
M = \alpha + a + F + bi + \beta{}i
   = \sum_{k=0}^4\langle M\rangle_k\,,\label{2.8}
\end{equation}
where $\alpha$ and $\beta$ are scalars, $a$ and
$b$ are vectors, and $F$ is a bivector. This is a
decomposition of $M$ into its $k$-vector parts,
with  $k = 0,1,2,3,4$, where $\langle\ldots \rangle_k$ means
``$k$-\textit{vector part}." Of course,
$\langle M\rangle_0=\alpha$, $\langle M\rangle_1=a$, $\langle M\rangle_2=F$,
$\langle M\rangle_3=bi$, $\langle M\rangle_4=\beta{}i$.
It is often convenient to drop the subscript on the scalar part, writing
$\langle M\rangle=\langle M\rangle_0.$

We say that a $k$-vector is even (odd) if the integer $k$ is even (odd). Accordingly, any multivector can be expressed as the sum of even and odd parts. A multivector is said to be ``\emph{even}" if its parts are even $k$-vectors. The even multivectors compose a subalgebra of the STA.
We will be using the fact that spinors  can be represented as even multivectors.

Computations are facilitated by the
operation of \textit{reversion}.
For $M$ in the expanded form (\ref{2.8}) the \textit{reverse}
$\tM$  can be defined by
\begin{equation}
\tM=\alpha + a - F - bi + \beta i\,.\label{2.9}
\end{equation}
For arbitrary multivectors $M$ and $N$
\begin{equation}
\widetilde{(MN)} = \tN\tM\,.\label{2.10}
\end{equation}

It is useful to extend the definitions of inner and outer products to multivectors of higher grade. Thus, for bivector $F$ and vector $a$ we can define inner and outer products
\begin{equation}
F\bdot a = \half(Fa -aF),\label{2.11}
\end{equation}
\begin{equation}
 F\wedge a = \half(Fa +aF),\label{2.11b}
\end{equation}
so that
\begin{equation}
Fa= F\bdot a+F\wedge a\,\label{2.12}
\end{equation}
expresses a decomposition of $Fa$ into vector and pseudovector parts.
For $F=b\wedge c$ it follows that
\begin{equation}
(b\wedge c)\bdot a= b(c\bdot a)-c(b\bdot a)\,.\label{2.13}
\end{equation}
Many other useful identities can be derived to facilitate coordinate-free computations. They will be introduced as needed throughout the paper.

Any fixed timelike vector such as $\{\gamma_0\}$ defines an inertial frame that determines a unique separation between space and time directions. Algebraically, this can be expressed as the \textit{``spacetime split"} of each vector $x$ designating a spacetime point (or event) into a \textit{time} component $ x\bdot\gamma_0=ct$ and a \textit{spatial position vector} $\mathbf{x}\equiv x\wedge \gamma_0$ as specified by the geometric product
\begin{equation}
x\gamma_0=ct +\mathbf{x}\,.\label{2.14}
\end{equation}
We call this a $ \gamma_{0} $-\textit{split} when it is important to specify the generating vector.
The resulting quantity $ct +\mathbf{x}$ is called a {\it paravector}
 
This ``split" maps a spacetime vector into the STA subalgebra of even multivectors where, by ``regrading,"  the bivector part can be identified as a spatial vector.
Accordingly, the even subalgebra is generated by a frame of ``spatial vectors"
 $\{\bsig_k \equiv \gamma_k\gamma_0; k= 1,2,3\}$, so that
\begin{equation}
\bsig_1\bsig_2\bsig_3= \gamma_0\gamma_1\gamma_2\gamma_3=i.\label{2.15}
\end{equation}
Obviously, this rendition of the STA even subalgebra is isomorphic to the Pauli algebra, though the Pauli algebra is not a subalgebra of the Dirac algebra because the matrix dimensions are different.

We use boldface letters exclusively to denote spatial vectors determined by a spacetime split. Spatial vectors generate a coordinate-free \textit{spatial geometric algebra} with the \textit{geometric product}
\begin{equation}
\a \b=\a \bdot \b + \a \wedge \b =\a \bdot \b + i\a\boldsymbol{\times} \b,\label{2.16}
\end{equation}
where $\a\boldsymbol{\times} \b=-i(\a\wedge \b)$ is the usual vector cross product.

For the even part $\langle M\rangle_+=Q$ of the  multivector $M$, a spacetime split gives us
\begin{equation}
Q = z  + F ,\label{2.16a}
\end{equation}
where  scalar and pseudoscalar parts combine in the form of a complex number
\begin{equation}
z = \alpha + i\beta ,\label{2.16b}
\end{equation}
and the bivector part splits into the form of a ``complex vector''
\begin{equation}
F = \E +i\B =-\tilde{F}. \label{2.17}
\end{equation}
Thus, the even subalgebra in STA has the formal structure of complex quaternions.
However, the geometric interpretation of the elements is decidedly different from the usual one assigned to quaternions.
Specifically, the bivector $i\B$ corresponds to a ``real vector'' in the quaternion literature. This difference stems from a failure to distinguish between vectors and bivectors dating back to Hamilton. For complex quaternions, it  reduces to failure to identify the   
imaginary unit $i$ as a pseudoscalar.  Geometric interpretation is crucial for application of  quaternions in physics.

Reversion in the subalgebra is defined by
\begin{equation}
Q^\dagger \equiv \gamma_0 \widetilde{Q}\gamma_0.  \label{2.17a}
\end{equation}
This is equivalent to ``complex conjugation'' of  quaternions.
In particular,
\begin{equation}
F^\dagger \equiv \gamma_0\tF\gamma_0 = \E -i\B,  \label{2.18}
\end{equation}
so that
\begin{equation}
\E=\half(F+F^\dagger), \qquad i\B=\half(F-F^\dagger). \label{2.19}
\end{equation}
Moreover,
\begin{equation}
FF^\dagger = \E^2 +\B^2+2\E\boldsymbol{\times} \B,  \label{2.20}
\end{equation}
\begin{equation}
F^2=F\bdot F+F\wedge F = \E^2 -\B^2+i\E\bdot \B,  \label{2.21}
\end{equation}
which are familiar expressions from electrodynamics.
the bivector $F$ is said to be \textit{simple} if
\begin{equation}
F\wedge F = 0 \quad \Leftrightarrow \quad \E\bdot \B=0,  \label{2.22}
\end{equation}
and that is said to be timelike, spacelike or null, respectively, when $F^2 = \E^2 -\B^2$ is positive, negative or zero.

Sometimes it is convenient to decompose the geometric product $FG$ into symmetric and antisymmetric parts 
\begin{equation}
FG = F\circ G + F\times G,  \label{2.22a}
\end{equation}
where the \textit{symmetric product} is defined by
\begin{equation}
F\circ G \equiv \half (FG + GF), \label{2.22b}
\end{equation}
and the \textit{commutator product} is defined by 
\begin{equation}
F\times G \equiv \half (FG - GF). \label{2.22c}
\end{equation}
In particular, for quaternions the symmetric product serves as a ``complex inner product,'' while the commutator product serves as an "outer product for complex vectors.'' Comparison with (\ref{2.16}) shows that for ``real vectors'' 
\begin{equation}
\a \circ \b=\a \bdot \b, \label{2.22d}
\end{equation}
and
\begin{equation}
\a \times  \b=\ \a \wedge \b =  i(\a\boldsymbol{\times} \b),\label{2.22e}
\end{equation}
Note that the cross product on the right is distinguished from the commutator product on the left of this equation by a boldface of the product symbol. Also, it should be understood that the equivalence of commutator and outer products in this equation does not generally obtain for arbitrary multivectors.
 
Concerning the spacetime split of products between even and odd multivectors, for a bivector $F = \E +i\B$ and spacteime vector $a$
with the split $a\gamma_0=a_0+\a$, we have
\begin{equation}
(F\bdot a)\gamma_0 = \E\bdot \a +a_0
\E+\a\boldsymbol{\times} \B . \label{2.23}
\end{equation}
This may be recognized as the form for a spacetime split of the classical Lorentz force. We will use it as a template for other spacetime splits later on.

Concerning differentiation, the derivative with respect to any multivector variable $M$ is denoted by $\partial_{M}$, so the derivative with respect to a vector variable $n$ is denoted by $\partial_{n}$. As the derivative with respect to a position vector $\x$ is especially important, we distinguish it with the special symbol   
\begin{equation}
\boldsymbol{\nabla}\equiv\partial_{\x}=\bsig_{k}\partial_{k},\label{2.28}
\end{equation}
in agreement with standard vector calculus.
Thus, for a relative vector field $\mathbf{A}=\mathbf{A}(\x)$ The identity (\ref{2.16}) gives us
\begin{equation}
\boldsymbol{\nabla}\mathbf{A}=\boldsymbol{\nabla}\bdot  \mathbf{A} + \boldsymbol{\nabla}\wedge \mathbf{A} =\boldsymbol{\nabla}\bdot\mathbf{A} + i\boldsymbol{\nabla}\boldsymbol{\times} \mathbf{A},\label{2.30}
\end{equation}
which relates the curl to the standard vector cross product.

For field theory, the derivative with respect to a spacetime point must be defined. Though that can be done in a completely coordinate-free way \cite{Hest03b}, for a rapid survey it is more expedient here to exploit the reader's prior knowledge about partial derivatives. 

For each spacetime point $x$ the reciprocal of a standard frame $\{\gamma^\mu\}$ determines a set of  ``rectangular coordinates"
$\{x^\mu\}$  given by
\begin{equation}
x^\mu = \gamma^\mu\bdot x\qquad\hbox{and}\qquad x = x^\mu\gamma_\mu\,.\label{2.24}
\end{equation}
In terms of these coordinates the derivative with respect to a spacetime point $x$ is an operator $\square $ defined by

\begin{equation}
\square\equiv\partial_x= \gamma^\mu\partial_\mu,\label{2.25}
\end{equation}
where
$\partial_\mu$ is  given by
\begin{equation}
\partial_\mu =\frac{\partial}{\partial{}x^\mu}=\gamma_\mu\bdot\square  \,.\label{2.26}
\end{equation}
The square of $\square$ is the usual d'Alembertian
\begin{equation}
\square^2=g^{\mu\nu}\partial_\mu
\partial_\nu\,\quad \hbox{where} \quad g^{\mu\nu}=\gamma^\mu\bdot\gamma^\nu.\label{2.27}
\end{equation}

The matrix representation of the {\it vector derivative} $\square$ will be
recognized as the so-called ``Dirac operator," originally
discovered by Dirac when seeking a ``square root"
of the d'Alembertian (\ref{2.27}) in order to find
a first order ``relativistically invariant" wave
equation for the electron. In STA however,
where the $\gamma^\mu$ are vectors rather than
matrices, it is clear that $\square$
is a vector operator, and we see that it is as significant in Maxwell's equations as in Dirac's.

The symbol $\nabla\equiv\partial_{x}$ is often used elsewhere \cite{Hest03b,Doran03} instead of $\square\equiv\partial_{x}$, but it has the disadvantage of confusability with $\boldsymbol{\nabla}\equiv\partial_{\x}$ in some contexts. Besides, 
the triangle is suggestive of three dimensions, while the $\square$ is suggestive of four.
That is why the $\square$ was adopted in the first book on STA
\cite{Hest66}, and earlier by Sommerfeld \cite{Sommerfeld52}
and Morse and Feshbach \cite{Morse53}.

Note that the symbol $\partial_{t}$ for the derivative with respect to a scalar variable $t$ denotes the standard partial derivative, though the coordinate index is used as the subscript in (\ref{2.26}).

In STA an electromagnetic field is represented by a
bivector-valued function $F = F(x)$ on spacetime. 
Since $\square$ is a vector operator the expansion (\ref{2.12}) applies, so  we can write
\begin{equation}
\square F = \square\bdot F + \square\wedge F\,,\label{2.29}
\end{equation}
where $\square\bdot F$ is the {\it divergence} of $F$ and $\square\wedge F$
is the {\it curl}.

Corresponding to the split of a spacetime point (\ref{2.14}), the spacetime split of the vector derivative $\square=\partial_x $ gives us a paravector derivative
\begin{equation}
\gamma_{0}\square=\partial_0+\boldsymbol{\nabla},\label{2.31}
\end{equation}
where $\partial_0=\gamma_{0}\bdot\square=c^{-1}\partial_{t}$.
Hence, for example, the {\it d'Alembertian} takes the familiar form
\begin{equation}
\square^{2}=\partial_{0}^{2}-\boldsymbol{\nabla}^{2},\label{2.33}
\end{equation}
and the divergence of the vector field $A=(c\varphi+\mathbf{A})\gamma_{0}$ splits to
\begin{equation}
\square\bdot A=\partial_0\,\varphi +\boldsymbol{\nabla}\bdot\mathbf{A}.\label{2.34}
\end{equation}
Finally, it is worth mentioning that to evaluate vector derivatives without resorting to coordinates, a few basic formulas are needed. For vector $n$ and bivector $F$, we shall have use for the following derivatives of linear functions:
\begin{equation}
\partial_{n}n=4,\quad\partial_{n}Fn=0,\quad\partial_{n}(n\bdot F)=2 F.\label{2.35}
\end{equation}

 \section{Anatomy of the Dirac wave function}\label{sec:II}

Considering the central role of Dirac's equation in the spectacular successes of quantum mechanics and QED, it seems indubitable that this compact equation embodies some deep truth about the nature of the electron, and perhaps elementary particles in general. However, success came with problems that called for action by the doctors of Quantum Mechanics. Soon after the initial success in explaining the hydrogen spectrum and the magical appearance of spin, it was discovered that the electron had an antiparticle twin, the positron, conjoined with it in the Dirac equation. Dirac introduced a surgical procedure called ``Hole theory" that suppressed the positron to keep it from interfering with the electron. Eventually, electron and positron were identified with positive and negative energy states and  separated by a procedure called ``second quantization." That has become the surgical procedure of choice in QED.
Here we take a new look at the anatomy of the Dirac equation to see what makes the electron tick. That leads to an alternative surgical procedure for separating the electron twins with new physical implications.

In terms of STA the Dirac equation can be written in the form
\begin{equation}
\gamma^\mu(\partial_\mu\Psi\i\hbar-\frac{e}{c}A_\mu\Psi) =m_{e}c\Psi\gamma_0\,,\label{4.1}
\end{equation}
where $m_{e}$ is electron mass and now we use $e=\pm |e|$ for the charge coupling constant, while the $A_\mu=A\bdot\gamma_\mu$ are components of the electromagnetic vector potential.
The symbol $\i$ denotes a unit bivector, which can be written in the following equivalent forms:
\begin{equation}
\i\equiv{}\gamma_2\gamma_1=i\gamma_3\gamma_0=i\bsig_3=\bsig_1\bsig_2\label{4.2}
\end{equation}
The notation $\i$ emphasizes that it plays the role of the unit imaginary that appears explicitly in matrix versions of the Dirac equation.

Let us refer to (\ref{4.1}) as the \emph{real Dirac equation} to distinguish it from the standard matrix version. It is well established that the two versions are mathematically isomorphic \cite{Hest69,Hest03b,Doran03}.
However, the real version reveals geometric structure in the Dirac theory that is so deeply hidden in the matrix version that it remains unrecognized by QED experts to this day. That fact is already evident in the identification of the imaginary unit $\i$ as a bivector. As we see below, this identification couples complex numbers in quantum mechanics inextricably to spin, with profound implications for physical interpretation. It is the first of several insights into geometric structure of Dirac theory that will guide us to a reformulation and new interpretation.

Employing the vector derivative puts the real Dirac equation in the coordinate-free form
\begin{equation}
\square\Psi\i\hbar-\frac{e }{c}A\Psi=m_{e}c\Psi\gamma_0\,,\label{4.3}
\end{equation}
where $A=A_\mu\gamma^\mu$ is the electromagnetic vector potential.
The spinor ``\emph{wave function}" $\Psi=\Psi(x)$ admits to the \emph{Lorentz invariant} decomposition
\begin{equation}
\Psi=\psi e^{i\beta/2} \quad \hbox{with}\quad \psi(x)=\rho^{\smallhalf} R(x) ,\label{4.4}
\end{equation}
where $\rho =\rho(x)$ and $\beta=\beta(x)$ are scalar-valued functions, and ``\emph{rotor}" $R= R(x) $ is normalized to
\begin{equation}
R\tR= \tR R=1.\label{4.5}
\end{equation}
The Lorentz invariant ``$\beta$-factor" in the general form (\ref{4.4}) for a ``\emph{Real Dirac spinor}" has been singled out for special consideration. As this factor is so deeply buried in matrix representations for spinors, its existence has not been generally recognized and its physical interpretation has remained problematic to this day. We shall see it as a candidate for corrective surgery on the Dirac wave function.

We shall also be considering singular solutions $\Psi_{\pm}$ of the Dirac equation (\ref{4.3}) called \emph{Majorana states} and defined by
\begin{equation}
\,\Psi_{\pm}=\Psi(1\pm\bsig_2)=\Psi\gamma_{\pm}\gamma_{0},\label{4.5a}
\end{equation}
where
\begin{equation}
\gamma_{\pm}=\gamma_{0}\pm \gamma_{2}.\label{4.5b}
\end{equation}
We shall see that STA reveals properties of these states that make them attractive candidates for distinct electron and positron states.

\subsection{Local Observables}

We begin physical interpretation of the Dirac wave function  with identification of ``\emph{local observables}."
At each spacetime point $x$, the rotor $R = R(x)$
determines a Lorentz rotation of a given fixed frame
of vectors $\{\gamma_\mu\}$ into a frame $\{e_\mu = e_\mu(x)\}$ given by
\begin{equation}
e_\mu=R\gamma_\mu\tR\,.\label{A.1}
\end{equation}
In other words, $R$ determines a unique frame field on spacetime.
Whence, the wave function determines four vector fields
\begin{equation}
\Psi\gamma_\mu \widetilde{\Psi}=\psi\gamma_\mu \tpsi=\rho e_\mu.\label{A.1b}
\end{equation}
Note that the $\beta$-factor has cancelled out of these expressions because the pseudoscalar $i$ anticommutes with the vectors $\gamma_\mu$.

It can be shown \cite{Hest69,Hest03b,Doran03} that two of the vector fields (\ref{A.1}) correspond to well known quantities in matrix Dirac theory.
The quantity
\begin{equation}
\psi\gamma_0 \tpsi=\rho v \quad\hbox{with}\quad v = R\gamma_{0}\tR= e_0.\label{A.2}
\end{equation}
is the \emph{Dirac current}. The Born interpretation identifies this as a ``\emph{probability current};" whence, $\rho$ is a \emph{probability density}. (We shall consider an alternative interpretation for $\rho$ later on.)
The quantity
\begin{equation}
s=\frac{\hbar}{2}R\gamma_{3}\tR =\frac{\hbar}{2}e_{3}\label{A.3}
\end{equation}
can be identified as the electron ``\emph{spin vector}," though it looks rather different than its matrix counterpart.

Physical interpretation of $e_{1}$ and $e_{2}$ is more subtle, as these vectors are not recognized in standard Dirac theory.
To clarify the matter, we decompose the rotor ${R}$ into the product
\begin{equation}
R=Ve^{-\i\varphi}.\label{A.4}
\end{equation}
Then
\begin{equation}
e_{1}=R\gamma_{1}\tR=e^{-I\varphi}a_{1}e^{I\varphi}=a_{1}e^{2I\varphi},\label{A.5}
\end{equation}
where
\begin{equation}
I\equiv R\,\i\tR=V\i\tV \quad \hbox{and}\quad a_{1}
=V\gamma_{1}\tV,\label{A.6}
\end{equation}
with an analogous equation for $e_{2}$. This exhibits the wave function \emph{phase} $\varphi$ as an angle of rotation in a spacelike plane with tangent bivector $I=I(x)$ at each spacetime point $x$. Moreover, the direction of that plane is determined by the \textit{spin bivector} defined by
\begin{equation}
S\equiv isv =\frac{\hbar}{2}R\,\i\tR
=\frac{\hbar}{2}I.\label{A.7} 
\end{equation}
Thus, we have a connection between spin and phase with the phase as an angle of rotation in the ``\textit{spin plane}."

In general, the Lorentz rotation (\ref{A.1}) has a unique decomposition into a spatial rotation followed by boost, which is generated by the rotor product \cite{Hest03b} 
\begin{eqnarray}
&R=VU \\
&\hbox{with}\quad U\gamma_{0}\tU=\gamma_{0}\quad\hbox{and}
\quad V=(v\gamma_{0})^{1/2}.\notag\label{A.7a} 
\end{eqnarray}
For simplicity, we often refer to rotors $V$ and $U$ by the same names ``boost'' and ``spatial rotation'' used for the Lorentz transformations they generate.

We can further decompose the rotor product into 
\begin{equation}
R=U_1 V_0 \tU_1 U=U_1 V_0 U_2 \label{A.7b}
\end{equation}
where 
\begin{equation}
 V_0=\exp{\{\alpha_{1}\bsig_{2}\}} =\cosh \alpha_{1} +\bsig_{2}\sinh \alpha_{1}
\label{A.7c} 
\end{equation}
is a boost in a fixed direction
$\bsig_{2}=\gamma_{2}\gamma_{0}$, while $U_1$ and $U_2$ are spatial rotations.

\subsection{Electron clock and chirality}

As the notion of an electron clock was central to de Broglie's seminal contribution to quantum mechanics \cite{Broglie87}, its relevance to interpretation of the Dirac equation deserves thorough investigation.  
The clock mechanism can be defined by considering a Dirac plane wave solution of the form (\ref{A.4}) with momentum $p$, wherein the phase has the specific form $\varphi=k\bdot x$. Then $ \square \varphi=k$, and the Dirac equation (\ref{4.3}) gives us
\begin{equation}
\hbar kRe^{i\beta/2}=m_{e}cRe^{i\beta/2}\gamma_{0},\label{A.8}
\end{equation}
which we solve for
\begin{equation}
k=\frac{m_{e}c}{\hbar}ve^{-i\beta}.\label{A.9}
\end{equation}
This has two solutions with opposite signs given by $\cos\beta=\pm 1$ and momentum $p= m_{e}c v=\pm \hbar k$,

Equation $v\bdot x=c\tau$  describes a propagating hyperplane with unit normal $v$, so (\ref{A.9}) gives
\begin{equation}
p\bdot x=m_{e}c^{2}\tau.\label{A.9a}
\end{equation}
Accordingly, the vector $e_{1} $ in (\ref{A.5}) rotates in (or on) the hyperplane with frequency
\begin{equation}
\omega_{e}\equiv\frac{2m_{e}c^{2}}{\hbar }=\pm 2\frac{d\varphi}{d\tau}.\label{A.10}
\end{equation}
The handedness is opposite for the two solutions
This will be recognized as the \emph{zitterbewegung} frequency of Schr\"odinger. It is precisely twice the \emph{de Broglie frequency} because the wave function phase angle is precisely half the rotation angle of the observables in (\ref{A.5}).
The sign of the phase specifies the sense of rotation, which is opposite for electron and positron. 

We can now give the vector $e_{1}$ a picturesque physical interpretation as \emph{the hand on de Broglie's electron clock}, with its rotation given by (\ref{A.5}). The \emph{face of the clock }is the bivector $I$ in (\ref{A.6}), and the reference point for an initial time on the clock face is given by the vector $a_{1}$. This description of the electron clock is completely general, as the equations hold for an arbitrary electron wave function. Indeed, equation (\ref{A.6}) shows that the electron clock can be described as an ``\emph{inertial clock},'' because it retains the mark of initial time even as interactions change the rotor $R$ and hence spin direction and the \emph{attitude } of the clock in spacetime.

Of course, interactions can change the clock frequency by changing the phase $\varphi$. Nevertheless, the free electron frequency remains as a reference standard for the electron clock. This suggests that we define the free electron clock period $\tau_{e}$ as the fundamental unit of electron time. Its empirical value, which I propose to call the \emph{zit}, is
\begin{equation}
\hbox{1 zit}=\tau_{e}=\frac{2\pi}{\omega_{e}}=\frac{h}{2m_{e}c^2}=
4.0466\boldsymbol{\times} 10^{-21} \mathrm{sec}\label{A.11}
\end{equation}
Approximately: 1 zit $\approx$ 4 zepto-sec;
 1 sec $\approx$ 1/4 zetta-zit.\\
Remarkably, direct measurement of the ``zit" may be possible with electron channeling experiments \cite{Hest10}.

The two signs in (\ref{A.10}) indicate clocks with opposite ``handedness" or \emph{chirality}, as we shall say. We identify the negative sign with an \emph{electron clock }and the the positive sign with a \emph{positron clock}. Indeed, in standard theory the two signs are interpretated as states with opposite energy and the negative energy state is identified with the positron. However, we have seen that the sign is actually determined by $\cos\beta=\pm 1$ without reference to a concept of energy. This suggests that we interpret $\beta$ as a ``chirality parameter." Be that as it may, we can see that the vector $e_{2}$ specifies the clock-face direction of motion for the clock hand $e_{1}$. Hence ``antiparticle conjugation" should be defined to reverse the direction of $e_{2}$ while keeping the direction of $e_{1}$ unchanged.

Finally, we note that the Born probability density has been set to $\rho=1$ on the propagating hyperplane, thus implying that all points on the hyperplane are equally probable positions $x_0$ for the electron at initial time $\tau_0$. However, for any initial position $x_0$,
the velocity $v=\dot{x}$ integrates to a unique position
\begin{equation}
x(\tau)=v\tau + x_0.\label{A.11a}
\end{equation}
Thus, the plane wave solution 
consists of an ensemble of equally probable particle paths composing a congruence (or \textit{fibration}) of  non-intersecting, timelike paths that sweep out (\textit{fibrate}) a region of spacetime.

\subsection{Electron clock with zitter}

There is another plane wave solution that has been largely overlooked in the literature. We simply switch (\ref{A.4}) into the form (with $\rho =1$)
\begin{equation}
\psi=e^{-\i\varphi}V_0,\label{A.11b}
\end{equation}
which is of type  (\ref{A.7b}) with constant $V_0$ given by  (\ref{A.7c}). This solves the Dirac equation with $\varphi=p\bdot x/\hbar$ and $p=m_e c\gamma_0$. To verify that:
\begin{equation}
\square\psi\i\hbar=-m_e c\gamma_0 i\bsig_3 \psi i \bsig_3 = m_e c \psi \gamma_0.\label{A.11bb}
\end{equation}
Note that $\gamma_0 i\bsig_3=i\gamma_3$ commutes with $\psi$, whereas $\gamma_0$ and $i\bsig_3$ do not. Generalization to a solution for arbitrary constant $p=m_e c V\gamma_0\widetilde{V}$ is obviously given by a boost to $\psi'=V\psi$.

Now, using (\ref{A.7c}) we can express the wave function (\ref{A.11b}) as the sum of positive and negative energy solutions:
\begin{equation}
\psi =\cosh \alpha_{1}e^{-\i k\bdot x} +\bsig_{2}\sinh \alpha_{1}e^{+\i k\bdot x} \equiv \psi_+ + \psi_- .
\label{A.11c} 
\end{equation}
The analog of hermitian conjugate in standard matrix Dirac algebra is defined by $\psi^\dagger =\gamma_0\tpsi \gamma_0.$
Whence, the velocity is given by
\begin{align}
v &= \psi\gamma_0 \tpsi = \psi\tpsi^\dagger\gamma_0 \notag \\ 
 &=\{|\psi_+|^2+|\psi_-|^2+\psi_+\psi_-^\dagger
 +\psi_-\psi_+^\dagger\} \gamma_0 \notag \\
 &=|\psi|^2\gamma_0 +2<\psi_-\psi_+^\dagger>\gamma_2 e^{\i 2\phi}. 
\label{A.11d} 
\end{align}
In agreement with \cite{Recami93}, this exhibits zitterbewegung as arising from interference between positive and negative energy states, as originally 
formulated by Schr\"odinger.
However, it also exhibits zitterbewegung as circulation of electron velocity in the spin plane.
I have coined the term \emph{zitter} to distinguish this interpretation of zitterbewegung from alternatives.

This result settles a long-standing controversy about the interpretation of zitterbewegung. To this day, studies of Dirac wave packets (e.g. \cite{Park12}) fail to recognize the connection of zitterbewegung to spin. Instead, it is identified as a high frequency interference effect, often attributed to interaction with the vacuum with a negative energy component $\psi_- $ presumed to express presence of positrons. On the contrary, in the zitter model here the ``negative energy'' term has nothing to do with positrons. Instead, it is a structural feature of electron motion involving  electron spin and phase.

We can associate our zitter plane wave with particle motion in the same way we did it for the plane wave in the preceding subsection. Without loss of generality, we can write $p=m_ec\gamma_0$, so $\varphi=p\bdot x/\hbar=\omega_e \tau /2$ defines a plane propagating in the direction of $p$ with proper time $\tau $.
Then (\ref{A.11b}) and (\ref{A.11d}) gives us a parametric equation for the particle velocity:

\begin{equation}
v(\tau)=e^{-\half\i \omega_e \tau}v_0\,
e^{\half\i \omega_e \tau}
= a\gamma_0 +b\,\gamma_2\, e^{\i \omega_e \tau},\label{A.11e}
\end{equation}
where $a$ and $b$ are constants, while $\omega_e$ is the free particle \emph{zitter frequency}. For $v=\dot{x}$, this integrates to
\begin{equation}
x(\tau)
= \gamma_0 ac\tau +b\lambda_e\,e_1 + x_0,\label{A.11f}
\end{equation}
where $\lambda_e=c/\omega_e$ and
\begin{equation}
e_1(\tau)=\gamma_1\, e^{\i \omega_e \tau},\label{A.11g}
\end{equation}
is the electron clock vector.

The particle path $x(\tau)$ specified by (\ref{A.11f}) is a timelike helix with pitch $b\lambda_e/a$.
Thus, the zitter plane wave solution 
consists of an ensemble of equally probable particle paths that fibrate a region of spacetime with
a congruence of  non-intersecting, timelike helices.

Though the circular frequency $\omega_e$ is constant, the circular speed increases with radius $b\lambda_e$ without reaching  the limiting case 
$\lambda_e \omega_e =c$ at the speed of light.
In that limit, $V_0 \rightarrow 1+\bsig_2$
in (\ref{A.7b}), and we get the Majorana wave function 
$\Psi_{+}$ defined in  (\ref{4.5a}), so
the velocity vector (\ref{A.11e}) becomes a null vector
\begin{align}
u(\tau)&=\Psi_{+}\gamma_0\widetilde{\Psi}_{+} \notag \\
&=e^{-\half\i \omega_e \tau}\gamma_{+}
e^{\half\i \omega_e \tau}
= \gamma_0 + \gamma_2\, e^{\i \omega_e \tau}.\label{A.11h}
\end{align}
In this case, zitter with electron clock is intrinsic to electron motion, whereas in the previous case described by (\ref{A.11e}) the zitter can vanish with $b=0$. 

Thus, we have three distinct kinds of free particle (plane wave) states: \textit{Kind A}, given by (\ref{A.4}), with no zitter; \textit{Kind B}, given by (\ref{A.11b}) and (\ref{A.11e}), with zitter velocity ranging between zero and the speed of light; and \textit{Kind C}, given by (\ref{A.11h}), with zitter velocity $\lambda_e \omega_e =c$.

\textit{Kind B} is related  to \textit{Kind A} by a unitary transformation.
For example, (\ref{A.11b}) is related to (\ref{A.4}) by
\begin{equation}
\gamma_1 (e^{-\i\varphi}V_0) \gamma_1^\dagger= V_0e^{-\i^\dagger\varphi},\label{A.11i}
\end{equation}
where $\gamma_1^\dagger=\gamma_1^*=-\gamma_1$
and the right side is interpreted as a positive energy factor with $\i$ replaced by $\i^\dagger$. It can be  generated by the continuous unitary  transformation
\begin{equation}
\psi \rightarrow W\psi W^{\dagger}, \quad \hbox{where}\quad
W=e^{\gamma_1 \, \alpha_0},\label{A.11j}
\end{equation}
which may be recognized as a Foldy-Wouthuysen (FW) transformation \cite{Foldy50}.

The FW transformation is commonly used to 
eliminate negative energy components in  electron wave functions, often because they are regarded as ``unphysical.''
Without going into arguments supporting this practice, the point here is that it suppresses the role of zitter in describing electron motion.

To ascertain what the Dirac equation can tell us about the physical significance of zitter, $\beta$ and electron clock, we study the properties of local observables thoroughly in the next section. This will help us address such questions as:
Is zitter an objectively real physical property of the electron? 
Should electron phase (de Broglie's clock) be regarded as a feature of electron zitter? What is the role of zitter in quantization? Of course, the answers will lead to more questions and speculation.

\subsection{Flow of Local Observables}

We turn now to a general analysis of conservation laws implied by the Dirac equation as a foundation for physical interpretation. To facilitate comparison with conventional Dirac theory, we first express the conservation laws in terms of the wave function. Then we peel them apart to reveal their structure in terms of local observables. 
 
A conservation law for the Dirac current $ \Psi\gamma_0\tilde{\Psi}=\rho v $ is easily derived from the Dirac equation (\ref{4.3}) and takes the form
\begin{equation}
\square\bdot(\rho v)=0.\label{B.1r}
\end{equation}
This can be interpreted as flow of a fluid with proper density $\rho$. Precisely what kind of fluid depends on the interpretation of other local observables, in particular, observables describing the flow of energy, momentum, charge and electromagnetic potential. Following a systematic approach in defining these observables within the Dirac theory, we shall discover hidden structure that has been generally overlooked.

The original formulation of the Dirac equation was based on interpreting
\begin{equation}
\underline{p}\,_{\mu}= \underline{i}\hbar\,\partial_{\mu}-\frac{e}{c}A_{\mu}\label{B.2r}
\end{equation}
as a gauge invariant energymomentum operator. The underbar notation here designates a linear operator. Specifically, the operator $ \underline{i} $  designates multiplication by the unit imaginary in the matrix version of Dirac theory, and right multiplication by the unit bivector 
$ \i= i\gamma_{3}\gamma_{0} $  in the STA version, as specified in 
\begin{equation}
\underline{p}\,_{\mu}\Psi = \hbar\,\partial_{\mu}\Psi i\gamma_{3}\gamma_{0}-\frac{e}{c}A_{\mu}\Psi. \label{B.2ra}
\end{equation}
Equivalence of operators in the matrix version to expressions in the present STA version is discussed in \cite{Hest03c}.

The energymomentum operator also led to the definition of an \textit{energy momentum tensor} $\underline{T}(n)$ with components 
\begin{equation}
T^{\mu \nu}=T^{\mu}\bdot\gamma^{\nu}
=\langle\gamma_{0}\widetilde{\Psi}
\gamma^{\mu}\underline{p}^{\nu}\Psi\rangle , \label{B.3r}
\end{equation}
where
\begin{equation}
T^{\mu}=
\underline{T}(\gamma^{\mu}).
\label{B.3rr}
\end{equation}
 The \textit{stress tensor} $ \underline{T}(n) $ \textit{is defined physically} as a vector-valued tensor field specifying, at each spacetime point, the energymomentum flux through a hypersurface with unit normal $n$. Its adjoint 
$ \overline{T}(n) $ can be defined by 
\begin{equation}
\gamma^{\mu}\bdot \overline{T}(\gamma^{\nu})= \underline{T}(\gamma^{\mu})\bdot\gamma^{\nu}=T^{\mu \nu}. \label{B.4r}
\end{equation}
Note the overbar notation $ \overline{T}(n) $ to indicate the adjoint of a linear function  $ \underline{T}(n) $ specified by an underbar. In this case the linear functions are vector-valued, but the same notation is used for bivector-valued linear functions below.

The Dirac equation (\ref{4.3}) can be derived from the Lagrangian
\begin{equation}
\mathcal{L}=\left\langle\hbar\,\square\Psi i\gamma_3\widetilde{\Psi}
-\frac{e }{c}A\Psi\gamma_0\widetilde{\Psi}- m_{e}c\Psi\widetilde{\Psi}
\right\rangle.\label{B.5r}
\end{equation}
As is well known, a major advantage of this approach is that conservation laws consistent with the equations of motion can be derived from symmetries of the Lagrangian. 
The most elegant and efficient way to do this is the  method of multivector differentiation introduced by Lasenby, Doran and Gull in \cite{Lasenby93}. In particular, from translation invariance of the Lagrangian they derived the stress tensor 
\begin{align}
\underline{T}(n) &=\gamma_{\mu}
\left\langle (\underline{p}^{\mu}\Psi) \gamma_{0}\widetilde{\Psi} n\right\rangle \notag\\
&=\gamma^{\mu}\left\langle (\hbar\partial_{\mu}\Psi i\gamma_{3}\gamma_{0}) \gamma_{0}\widetilde{\Psi} n\right\rangle -\frac{e}{c}A\rho\,
(v\bdot n).\label{B.6r}
\end{align}
This is equivalent to the stress tensor most commonly employed in Dirac theory. 

However, when Lasenby, Doran and Gull generalized their method in a ground breaking paper on Gauge Theory Gravity \cite{LDG98}, translation invariance gave instead the adjoint stress tensor 
\begin{equation}
\overline{T}(n) 
=\left\langle \hbar( n\bdot\square\Psi) i\gamma_{3}\widetilde{\Psi} \right\rangle_{1} -\frac{e}{c}(A\bdot n)\rho\, v.\label{B.7r}
\end{equation}
This raises a question as to which stress tensor is correct for the electron: $ \underline{T}(n) $ or $ \overline{T}(n) $? We leave the question open for the time being while we examine and compare their properties. 

The dynamics of flow is determined by the divergence of the stress tensor: 
\begin{align}
\grave{{\underline{T}}}(\grave{\square})&=\partial_{\mu}\underline{T}(\gamma^{\mu})=\partial_{\mu}T^{\mu}\notag\\&=\left\langle \hbar( \square^{2}\Psi) i\gamma_{3}\widetilde{\Psi} \right\rangle_{1} -\frac{e}{c}\partial_{\mu}(\rho\, vA^{\mu}).\label{B.8r}
\end{align}
We need the Dirac equation (\ref{4.3}) to evaluate this. Since
\begin{equation}
\left\langle \partial_{\mu}\Psi i\gamma_{3}\,\partial^{\mu}\widetilde{\Psi}\right\rangle_{1} =0,\label{B.9r}
\end{equation}
we have
\begin{align}
\left\langle \hbar( \square^{2}\Psi) i\gamma_{3}\widetilde{\Psi} \right\rangle_{1}&=
\frac{\hbar}{2}[ \square^{2}\Psi i\gamma_{3}\widetilde{\Psi}
- \Psi i\gamma_{3}\square^{2}\widetilde{\Psi}] \notag\\
&= \rho\frac{e}{c}(\square\wedge A)\bdot v+\frac{e}{c}\partial_{\mu}(\rho\, vA^{\mu}).\label{B.10r}
\end{align}
Whence
\begin{equation}
\grave{\underline{T}}(\grave{\square})=\partial_{\mu}T^{\mu}= \frac{e }{c}F\bdot(\rho v)\equiv \rho f,
\label{B.11r}
\end{equation}
where $ F=\square\wedge A $ is an external electromagnetic field.
This has precisely the form for the Lorentz force on a classical charged fluid, and it  supports the interpretation of the Dirac current $ e\rho v $ as a charge current.

A conservation law for angular momentum can be derived from invariance of the Lagrangian (\ref{B.5r}) under Lorentz rotations \cite{Lasenby93}, but we derive it directly from properties of the stress tensor, as it makes structure more explicit.
From (\ref{B.11r}) we get
\begin{equation}
\partial_{\mu}(T^{\mu}\wedge x)=T^{\mu}\wedge\gamma_{\mu}+\rho f\wedge x.
\label{B.12r}
\end{equation}
To see how this equation gives us angular momentum conservation, we need to analyze the first term on the right. In doing so we find other interesting results as byproducts.

First, note that 
\begin{equation}
\gamma^{\mu}\left\langle \hbar(\partial_{\mu}\Psi) i\gamma_{3}\widetilde{\Psi} \right\rangle_{1} 
=\hbar(\square\Psi) i\gamma_{3}\widetilde{\Psi}+\square(i\rho s).\label{B.13r}
\end{equation}
Then, combine this with the Dirac equation (\ref{4.3}) in the form
\begin{equation}
\hbar(\square\Psi) i\gamma_{3}\widetilde{\Psi}
=m_{e}c\rho e^{i\beta}+\frac{e}{c}A\rho v\label{B.14r}
\end{equation}
to get
\begin{equation}
\partial_{n}\overline{T}(n)=\gamma_{\mu}\overline{T}(\gamma^{\mu})
=\square(i\rho s)+m_{e}c\rho e^{i\beta}.
\label{B.15r}
\end{equation}
The scalar part of this equation gives us the trace of the stress tensor:
\begin{equation}
Tr(\overline{T})=\partial_{n}\bdot \overline{T}(n)=m_{e}c\rho \cos \beta,
\label{B.16r}
\end{equation}
and the pseudoscalar part gives us
\begin{equation}
\square\bdot (\rho s)=m_{e}c\rho \sin \beta.
\label{B.17r}
\end{equation}
This displays a peculiar relation of $ \beta $ to mass and spin of questionable physical significance. However, $ \beta $ plays no role in the bivector part of (\ref{B.15r}), which gives us 
\begin{equation}
\gamma_{\mu}\wedge \overline{T}(\gamma^{\mu})=\underline{T}(\gamma^{\mu})\wedge\gamma^{\mu}=\square\bdot (i\rho s)=
\partial_{\mu} S^{\mu},\label{B.18r}
\end{equation}
where
\begin{equation}
S^{\mu}=\underline{S}(\gamma^{\mu})= \gamma^{\mu}\bdot(i\rho s)=\rho i(s\wedge\gamma^{\mu}).
\label{B.19r}
\end{equation}
is identified as a bivector-valued spin flux tensor.

Equation (\ref{B.18r}) gives us an explicit relation between the stress tensor and its adjoint:
\begin{equation}
\overline{T}(n)-\underline{T}(n)=n\bdot(\gamma_{\mu}\wedge T^{\mu})=(n\wedge\square)\bdot (i\rho s).\label{B.20r}
\end{equation}
And inserting this into (\ref{B.11r}), we find that the divergence of the stress tensor is equal to the divergence of its adjoint: 
\begin{equation}
\partial_{\mu}\overline{T}(\gamma^{\mu})=\partial_{\mu}\underline{T}(\gamma^{\mu})= \frac{e }{c}F\bdot(\rho v)= \rho f.
\label{B.21r}
\end{equation}
This equivalent divergence of the stress tensor and its adjoint has been overlooked in the literature. Let's consider the difference more closely.

The flux of momentum in the direction of the Dirac current is especially significant, because that is the flow direction of electric charge. Accordingly, we define a momentum density $ \rho p $ along this flow by
\begin{equation}
\rho p\equiv \underline{T}(v).
\label{B.22r}
\end{equation}
The adjoint determines a ``\textit{conjugate momentum}" density  $ \rho p_{c} $ defined by
\begin{equation}
\rho p_{c}\equiv \overline{T}(v).
\label{B.23r}
\end{equation}
We will be looking to ascertain the physical difference between these two kinds of momenta. First we note a small difference in angular momentum.

Returning now to the question of angular momentum conservation,
inserting (\ref{B.18r}) into (\ref{B.12r}), we get the desired conservation law:
\begin{equation}
\underline{\grave{J}}(\grave{\square})=\partial_{\mu}J^{\mu}=\rho f\wedge x.
\label{B.23ar}
\end{equation}
where the total angular momentum tensor flux is a bivector-valued tensor with orbital and spin parts defined by
\begin{equation}
J^{\mu}=\underline{J}(\gamma^{\mu})=T^{\mu}\wedge x+S^{\mu}.
\label{B.23br}
\end{equation} 
Accordingly, the angular momentum flux along the Dirac current is given by 
\begin{equation}
\underline{J}(v)=\rho( p\wedge x+S),
\label{B.23cr}
\end{equation}
where $ \underline{S}(v) = \rho S $ confirms our earlier identification of $ S=isv $ as a spin bivector.

Alternatively, we can define a ``conjugate" angular momentum tensor  
\begin{equation}
J_{c}^{\mu}=\overline{T}(\gamma^{\mu})\wedge x -S^{\mu},
\label{B.24ar}
\end{equation} 
which by the same argument yields the conservation law
\begin{equation}
\partial_{\mu}J_{c}^{\mu}=\rho f\wedge x,
\label{B.24br}
\end{equation}
But the conjugate angular momentum flow has a spin of opposite sign:
\begin{equation}
\overline{J}_{c}(v)=\rho( p_{c}\wedge x-S).
\label{B.24cr}
\end{equation}
This sign difference can be interpreted geometrically as an opposite orientation of spin $ S $ to velocity $v$ or momenta $p$ and $p_{c}$.  
To probe the difference between the momenta $ p $ and  $ p_{c} $ more deeply, we express them as explicit functions of local observables.

The dynamics of the local observables $e_{\mu}=R\gamma_\mu\tR $ is determined by the linear bivector-valued function
\begin{equation}
\Omega_{\mu}=\underline{\Omega}(\gamma_{\mu})\equiv 2(\partial_{\mu}R)\tilde{R}.\label{B.24r}
\end{equation}
Thus,
\begin{equation}
\partial_{\nu}e_\mu=\Omega_{\nu}\bdot e_\mu.\label{B.25r}
\end{equation}
In particular, the derivatives of the velocity and spin vectors are 
\begin{equation}
\partial_{\nu}v=\Omega_{\nu}\bdot v \quad \hbox{and} \quad
\partial_{\nu}s=\Omega_{\nu}\bdot s ,\label{B.26r}
\end{equation}
while the derivative of the spin bivector $ S= isv$ is given by the commutator product:
\begin{equation}
\partial_{\mu}S=\Omega_{\mu}\times S.\label{B.27r}
\end{equation}

Now, with the wave function in the form
\begin{equation}
\Psi=\psi e^{i\beta/2} =Re^{(\alpha + i\beta)/2},\label{B.28r}
\end{equation} 
its derivatives can be related to observables by 
\begin{equation}
\hbar (\partial_{\mu}\Psi )i\gamma_3\widetilde{\Psi}=\rho
[(i\partial_{\mu}\alpha + \partial_{\mu}\beta)s+\Omega_{\mu}Sv],\label{B.29r}
\end{equation}
with the product expansion 
\begin{equation}
\Omega_{\mu}S=P_{\mu}+\partial_{\mu}S+iq_{\mu}\label{B.30r}
\end{equation} 
where we have identified components $ P_{\mu}=\gamma_{\mu}\bdot P $ of the \textit{``canonical momentum vector"} $ P $ defined by
\begin{equation}
P_{\mu}=\Omega_{\mu}\bdot S
=\frac{\hbar}{2}e_{1}\bdot \partial_{\mu}e_{2}=-\frac{\hbar}{2}e_{2}\bdot \partial_{\mu}e_{1},\label{B.31r}
\end{equation}
and the pseudoscalar part is given by
\begin{equation}
iq_{\mu}=\Omega_{\mu}\wedge S=i\Omega_{\mu}\bdot (sv)=i(\partial_{\mu}s)\bdot v.\label{B.32r}
\end{equation} 
Finally, by inserting (\ref{B.29r}) into  (\ref{B.7r}) we get the components of the stress tensor in the transparent form  
\begin{equation}
T_{\mu\nu}=\rho[v_{\mu}(P_{\nu}-\frac{e }{c}A_{\nu})+(v\wedge\gamma_{\mu})\bdot\partial_{\nu}S-
s_{\mu}\partial_{\nu}\beta] .\label{B.33r} 
\end{equation}
This gives us an informative expression for the conjugate momentum:
\begin{equation}
\hspace*{-.1in}\overline{T}(v)=\rho \{[(P-\frac{e }{c}A )\bdot v]v+ \dot{S}\bdot v - s\dot{\beta}\}=\rho p_{c}.\label{B.35r}
\end{equation}
The first two terms in this expression for momentum flow along a streamline of the Dirac current make perfect physical sense. 
Note that the factor $ (P-\frac{e }{c}A )\bdot v $ serves as a gauge invariant variable mass determined by the frequency of the electron clock, which is specified by 
\begin{equation}
P\cdot v=\Omega_{v}\bdot S
=\frac{\hbar}{2}e_{1}\bdot \dot{e}_{2}=-\frac{\hbar}{2}e_{2}\bdot \dot{e}_{1},\label{B.36r}
\end{equation}
where
\begin{equation}
\dot{e}_\mu=v\bdot\square\, e_\mu =\Omega_{v}\bdot e_\mu
\quad \hbox{and} \quad \Omega_{v}=\underline{\Omega}(v).\label{B.37r}
\end{equation}
The second term $ \dot{S}\bdot v=\dot{v}\bdot S$ in (\ref{B.35r}) specifies a contribution of spin to linear momentum due to acceleration. 
However, a physical interpretation for the last term involving the directional derivative $ \dot{\beta}=v\bdot \square \beta $ remains problematic.

From the stress energy components (\ref{B.33r}), we also get a remarkably simple expression for the  momentum density:
\begin{equation}
\underline{T}(v)=\rho (P-\frac{e }{c}A )=\rho p.\label{B.38r}
\end{equation}
And for flux in the spin direction we get:
\begin{equation}
\underline{T}(\hat{s})=\frac{\hbar}{2}\rho \,\square \beta.\label{B.38rs}
\end{equation}
Combining (\ref{B.38r}) with (\ref{B.35r}) we get
\begin{equation}
p_{c}= (p\bdot v)v+ \dot{S}\bdot v
 - s\dot{\beta}.\label{B.38rr}
\end{equation}
As we shall see, it is especially important to note that in these equations both $p$ and $p_{c}$ are defined independently of $\rho$, and physical interpretation of the strange parameter $\beta$ appears to be tied up with spin. 

Having thus identified the canonical momentum $P$ as a local observable, we can express the Dirac equation as a constitutive equation relating observables.
Thus, from (\ref{B.29r}) and  (\ref{B.30r}) we derive the expression
\begin{equation}
\hbar (\square\Psi )i\gamma_3\widetilde{\Psi}=[\rho P+
[\square(\rho e^{i\beta}S)]e^{-i\beta}]v,\label{B.39r}
\end{equation}
which we insert into the Dirac equation (\ref{B.14r}) to get it in the form
\begin{equation}
\rho (P-\frac{e }{c}A )e^{i\beta}=m_{e}c\rho v-\square(\rho e^{i\beta}S)\label{B.40r}
\end{equation}
Its vector part is a constitutive equation involving the Dirac current:
\begin{equation}
\rho (P-\frac{e }{c}A )\cos{\beta}=m_{e}c\rho v-\square\bdot (\rho e^{i\beta}S).\label{B.41r}
\end{equation}
The right side of this equation has vanishing divergence, and we identify it
as the well known \textit{Gordon current}. 
Unlike the vector part, the trivector part of (\ref{B.40r}) does not have any evident physical meaning, though it does serve as a constraint among the variables.


This completes our exact reformulation of Dirac Theory in terms of local observables. We have found clear physical interpretations for all components of the Dirac wave function except the parameter $\beta$. The strangeness of $\beta$
is most explicit in equation (\ref{B.41r}) for the Gordon current, where the factor $e^{i\beta}$  generates a duality rotation without obvious physical significance. And that equation implies the conservation law
\begin{equation}
\square\bdot[\rho (P-\frac{e }{c}A )\cos{\beta}]=\square\bdot(\rho p \cos\beta)=0,\label{B.41rq}
\end{equation}
where again the role of $\beta$ is problematic.

The Gordon current can be regarded as a reformulation of the Dirac equation in terms of local observables, as our derivation of (\ref{B.41r}) shows. For this reason, it plays a fundamental role in our analysis of alternative physical interpretations in subsequent sections. 
But first we try to make some sense of $\beta$.

\subsection{Problems with $ \beta $}

We begin our study of $\beta$ by reformulating the Dirac Lagrangian (\ref{B.5r}) with $\Psi =\psi e^{i\beta/2}$ to make the role of $ \beta $ explicit and then to relate it to the explicit role of other observables:
\begin{align}
\mathcal{L}&=\left\langle\hbar\,\square\psi i\gamma_3\widetilde{\psi}
-\frac{e }{c}A\psi\gamma_0\widetilde{\psi}- m_{e}c\rho\cos\beta
-\rho s\square\beta
\right\rangle \notag\\
&\qquad=\rho(P-\frac{e}{c}A)\bdot v +(v\wedge\square)\bdot (\rho S) \notag\\
&\qquad\qquad- m_{e}c\rho\cos\beta
-\rho s\bdot\square\beta .\label{B.42r}
\end{align}
The mass term $ <m_{e}c\Psi\widetilde{\Psi}>\,=m_{e}c\rho \cos\beta $ has always been problematic in QED. Indeed, it has been eliminated from the Standard Model, which aims to derive the mass from fundamental theory. 

When the $\beta$-factor $e^{i\beta}$ is constant, (\ref{4.4})  can be used to factor it out of the Dirac equation (\ref{4.3}) to exhibit its role explicitly:
\begin{equation}
\{\hbar\square\psi i\gamma_{3}\gamma_{0}-\frac{e }{c}A\psi\}\tpsi e^{i\beta}
=m_{e}c\psi\gamma_0\tpsi=m_{e}c\rho v. \label{C.4}
\end{equation}
Note that setting $e^{i\beta}=-1$ amounts to reversing orientation of the bivector $\i=i\gamma_{3}\gamma_{0}$ that generates rotations in the phase plane along with reversing the sign of the charge, as required for antiparticle conjugation according to the chirality hypothesis. Accordingly, the Dirac equation  is resolved into separate equations for electron and positron.

We have seen how plane wave solutions of the Dirac equation suggest that the  $\beta$ distinguishes particle from antiparticle states. Let me call that suggestion the ``\emph{chirality hypothesis}." Some credence to this hypothesis is given by the fact that unitary spinors $R$ and $Ri$ are distinct spin representations of the Lorentz group, so it is natural to associate them with distinct particles. However, the Dirac spinor $Re^{i\beta/2}$ is a continuous connection between both representations, suggesting that $\beta$ parametrizes an admixture of particle/antiparticle states.

After I discovered that $\cos\beta=\pm 1$ solves the problem of negative energies for plane waves and thereby separates electron and positron plane wave states \cite{Hest69}, I set about studying the physical significance of $\beta$ in the general case. I got great help from my graduate student Richard Gurtler, who thoroughly examined the behavior of $\beta$ in the \emph{Darwin solutions} of the Dirac equation for Hydrogen \cite{Gurtler72}. The results do not seem to support the chirality hypothesis, for the parameter $\beta$ varies with position in peculiar ways. The values $\cos\beta=\pm 1$ appear only in the azimuthal plane, which suggests that 2d solutions might satisfy the chirality hypothesis, but the chirality jumps in sign across nodes in the plane in an unphysical way. At about the same time I began a systematic study of local observables in Dirac theory \cite{Hest73} and their roles in Pauli and Schr\"odinger theories \cite{Hest75a,Hest75b,Hest79} , but  I was unable to make sense of the peculiar behavior I found for $\beta$.

This problem of interpreting $\beta$ has never been recognized in standard QED. Indeed, it is commonly claimed that second quantization solves Dirac's problem of negative energies. My suspicion is that $\cos\beta=\pm 1$ has been tacitly assumed in QED when it begins by quantizing plane wave states. Consistency of that procedure with the Darwin solutions has never been proved to my knowledge. It seems that a perceived need for such a proof is avoided by claiming that the Darwin case is concerned with one-particle quantum mechanics, whereas QED is a many-particle theory.

To nail down the interpretation of $\beta$, we need to study its role in specific solutions of the Dirac equation. A step in that direction will be taken in the next Section. In the meantime, it is worth looking at simplifications when $\cos \beta= \pm 1 $.

With $\cos \beta=1 $ the Dirac equation (\ref{B.40r}) reduces to 
\begin{equation}
 \rho (P-\frac{e }{c}A )=m_{e}c\rho v-\square(\rho S).\label{C.5}
\end{equation}
To reiterate an important remark: \textit{the physical content of this equation resides \textit{entirely} in its vector part.} The trivector part 
\begin{equation}
\square\wedge (\rho S)=\square\bdot (\rho sv)i=0.\label{C.6}
\end{equation}
appears to be a constraint on density flow
beyond the conservation laws
\begin{equation}
\square\bdot(\rho v)=0\quad\hbox{and}\quad\square\bdot(\rho s)=0.\label{C.6a}
\end{equation}
Obviously, these are kinematic constraints on local observables independent of external interactions.

Comparison of (\ref{C.5}) with (\ref{B.38r}) shows that the vector part equates the momentum density to the Gordon current: 
\begin{equation}
\rho p =\rho(P-\frac{e }{c}A)=m_{e}c\rho v-\square\bdot(\rho S).\label{C.10}
\end{equation}
The conservation law for the Dirac current and the identity   
\begin{equation}
\square\bdot [\square\bdot(\rho S)]=(\square\wedge \square)\bdot(\rho S)=0 \notag
\end{equation} 
then gives us the Gordon conservation law 
\begin{equation}
\square\bdot(\rho p)=0.\label{C.11}
\end{equation}
Further, note that the momentum $ p=P-(e/c)A $ 
is independent of $\rho$.
Its curl has the strikingly simple form
\begin{equation}
\square\wedge p=-\frac{e }{c} F+\square\wedge P,\label{C.45}
\end{equation}
where $ F=\square\wedge A $ is the external electromagnetic field.
Dotting with velocity $v$ we get
\begin{equation}
\dot{p}=\frac{e }{c} F\bdot v+v\bdot (\square\wedge P) +f_{s}(v),\label{C.46}
\end{equation}
which looks like an equation of motion with Lorentz force, though functional dependence of $P$ and $ f_{s}(v)\equiv\grave{\square} \,\grave{p}\bdot v $ remain to be determined.

Now, to relate the Gordon current to the conjugate momentum, we note that  $v\bdot S=0$ implies
\begin{equation}
 (v\wedge\square)\bdot(\rho S)=\rho(\square\wedge v)\bdot S. \notag
\end{equation} 
Hence the momentum equation (\ref{C.10}) gives us
a density-free expression for generalized mass:
\begin{equation}
v\bdot p =m_{e}c+S\bdot(\square\wedge v).
\label{C.11b}
\end{equation}
Whence (\ref{B.38r}) gives us an explcit expression for the conjugate momentum:
\begin{equation}
p_{c}= [m_{e}c+S\bdot(\square\wedge v)]v+ \dot{S}\bdot v.\label{C.11c}
\end{equation}
To understand this better, we examine what the Dirac equation says about  the velocity curl.


Since the Dirac current $ \rho v $ is conserved, it determines a congruence of velocity streamlines, which can be regarded as timelike paths. We are interested in  how local observables evolve along each path. The evolution of the \textit{comoving frame } {$ e_\mu=R\gamma_\mu\tR\ $} along a path is determined by the equation of motion
\begin{equation}
\dot{e}_\mu=v\bdot\square\, {e}_\mu=\Omega\bdot e_\mu,\label{C.29}
\end{equation}
with \textit{proper angular velocity}
\begin{equation}
\Omega\equiv v^{\mu} \Omega_{\mu}
=2(v\bdot\square R)\tR.\label{C.30}
\end{equation}
However, the proper definition of the time  derivative for rotor $R$ is tricky. We must look to the Dirac equation to see how to do that and to express $\Omega$ in terms of observables.
The Dirac equation (\ref{C.4}) with $\cos \beta=1$ can be put in the form
\begin{align}
 2(\square\psi)\psi^{-1}&=2(\square R)\tR+\square\ln\rho \notag\\
&=(m_{e}cv+\frac{e }{c}A)S^{-1}.\,\, \label{C.31}
\end{align}
We can eliminate $ R $  with the identity
\begin{equation}
(\square R)\tR v +v(\square R)\tR=\square v+2(v\bdot\square R)\tR,\label{C.32}
\end{equation}
and then separate scalar and bivector parts.

From the scalar part  of (\ref{C.32}) we get the familiar conservation law for the Dirac current:
 \begin{equation}
\square\bdot(\rho v)=0.\label{C.32a}
 \end{equation} 
But note now that this law plays an essential role in defining an invariant time derivative for local observables as the directional derivative along a particle path. Thus, 
\begin{equation}
v\bdot\square \ln \rho =\dot{\rho}/\rho=-\square\bdot v.\label{C.32b}
\end{equation}
To be invariant, the time derivative of rotor $R$ must be defined differently.

From the bivector part of (\ref{C.32}) we get
\begin{equation}
\square\wedge v+2(v\bdot\square R)\tR
=(m_{e}c+\frac{e }{c}A\bdot v)S^{-1} .
\label{C.33}
\end{equation}
In other words,
\begin{equation}
\square\wedge v+\Omega
=(m_{e}c+\frac{e }{c}A\bdot v)S^{-1} .
\label{C.33a}
\end{equation}
This is the expression for $\Omega$ that we were looking for. It is eminently reasonable that rotation along Dirac streamlines be determined by the curl of velocity. And the term generating rotation in the spin plane specifies time evolution of the electron clock, including the effect of external potentials. 
Thus, for constant velocity it reduces to
\begin{equation}
c\Omega=2c(v\bdot\square R)\tR
=m_{e}c^{2}S^{-1}=-\omega_{e}I,\label{C.32cc}
\end{equation}
where $\omega_{e}=2m_{e}c^{2}/\hbar$ is the familiar free particle zitter frequency and
$ I=e_{2}e_{1}=2S/\hbar $ is the unit spin bivector.

For the comoving frame (\ref{C.33}) gives us
\begin{equation}
\dot v=\Omega\bdot v =v\bdot(\square\wedge v)=\Omega\bdot v,\label{C.34}
\end{equation}
\begin{equation}
\dot s=\Omega\bdot s =s\bdot(\square\wedge v)=\Omega\bdot s.\label{C.35}
\end{equation}
and
\begin{align}
\Omega\bdot I&= e_{1}\bdot\Omega\bdot{e}_{2}
=e_{1}\bdot\dot{e}_{2}=-e_{2}\bdot\dot{e}_{1}
\notag\\
&=-I\bdot(\square\wedge v)
+(\omega_{e}+eA\bdot v)/c. \label{C.36}
\end{align}
This last result describes more explicitly the evolution of electron phase or, if you will, time on the electron clock.

This completes our analysis of geometric structure in Dirac Theory. In the following we apply it to physical interpretation, with special emphasis on particle properties and the roles of phase (in zitter),  ``density'' $\rho$ and the overlooked parameter $\beta.$

\section{Pilot Wave Theory with the Dirac Equation }\label{sec:IV}

Let me coin the name \textit{Born--Dirac} for standard Dirac theory with the   \textit{Born rule} for interpreting the Dirac wave function as a probability amplitude. 

The Born rule was initially adopted for Schr\"odinger theory and subsequently extended to Dirac theory without much discussion --- in fact, without even establishing the correct relation between Dirac and Schr\"odinger wave functions.
The latter is supposed to describe a particle without spin. However, a correct derivation from the Dirac equation
\cite{Hest75b,Hest79}
implies instead that the Schr\"odinger equation describes an electron in a spin eigenstate, and its imaginary unit must be identified with the spin bivector $\i\hbar = 2 i\s$ .

Subsequently, physical interpretation of 
Schr\"odinger theory has been hotly debated, while, ironically, relevant implications of the more precise Dirac theory have been overlooked.
To correct this deficiency, 
our first task here is to update Born-Dirac theory
with recent insights on interpretation of 
Schr\"odinger theory. Then we can consider enhancements from our study of local observables in Dirac theory.

After decades of debate and clarifications, it  seems  safe to declare  that de Broglie--Bohm ``\textit{Pilot Wave}'' theory is well established as a viable interpretation of quantum mechanics, though that may still be a minority opinion among physicists.
Current accounts suitable for our purposes are given in \cite{Norsen13,Holland93}.
The point to be emphasized here can be regarded as a refinement of the Born rule, which says the wave function for a single electron specifies its probable position at a given time. The \textit{Pilot Wave rule} extends that to regarding the wave function as specifying an ensemble of possible particle paths, with the electron traversing exactly one of those paths, but with a certain probability for each path. So to speak, the wave function serves to guide the electron along a definite path, but with a specified probability. Hence the name \textit{``pilot wave''} for the wave function. In his ``theory of the double solution,'' de Broglie argued for a physical mechanism to select precisely one of those paths,
but that alternative is not available in conventional Pilot Wave theory. Instead, path selection is said to require an act of observation, which continues to be a subject of contentious debate and will not be discussed here.

Strictly speaking the \textit{Pilot-Wave rule} requires only an assignment of particle paths to interpret the wave function; whence, $\rho(\x,t)$ can be interpreted as a \textit{ density of paths.}  However, for agreement with the Born rule it allows
assignment of probabilities to the wave function in its initial conditions, which then propagate to probabilities at any subsequent time. 
Accordingly, these probabilities should not be interpreted as expressions of randomness inherent in Nature as commonly claimed for Schr\"odinger theory.
Rather, consistent with its realist perspective, Pilot Wave theory regards probabilities in quantum mechanics  as expressing limitations in  knowledge of specific particle states (0r paths).
This viewpoint is best described by  Bayesian probability theory, as most trenchantly expounded by Jaynes  \cite{Jaynes03}.
Accordingly, we regard the Born-Dirac wave function as specifying Bayesian conditional probabilities for electron paths.

The Schr\"odinger wave function in Pilot Wave theory is a many particle wave function.  Here we confine attention to the single particle theory, and we review some well known specifics  \cite{Holland93} to focus on crucial points.

With wave function $\psi =\rho^{1/2} e^{ S/\i\hbar}$,
Schr\"odinger's equation can be split into a pair of coupled equations for real functions $\rho=\rho(\x,t)$ and  $S=S(\x,t)$ with scalar potential $V=V(\x)$:   
\begin{equation}
\partial_t S+\frac{(\boldsymbol{\nabla} S)^2}{2m}
-\frac{\hbar^2}{2m}\frac{\boldsymbol{\nabla}^2{\rho^{1/2}}}{\rho^{1/2}} +V=0,
  \label{4.10a} 
\end{equation}
\begin{equation}
\partial_t \rho + \boldsymbol{\nabla}\cdot\left( \frac{\rho\boldsymbol{\nabla} S}{2m}\right) =0.
   \label{4.10b} 
 \end{equation}
Equation (\ref{4.10a}) can be written
\begin{equation}
( \partial_t+\frac{1}{m}(\boldsymbol{\nabla} S)\cdot\boldsymbol{\nabla} ) \boldsymbol{\nabla} S = -\boldsymbol{\nabla} (V+Q),  \label{4.10c} 
\end{equation}
where
\begin{equation}
Q=Q(\x,t)=\frac{\hbar^2}{2m}
\frac{\boldsymbol{\nabla}^2{\rho^{1/2}}}{\rho^{1/2}} .
  \label{4.10d} 
\end{equation}
Identifying 
\begin{equation}
m^{-1}\boldsymbol{\nabla} S=\v=\dot{\x} 
  \label{4.10dd}  
\end{equation}
as the velocity of a curve  $\x(t)$ normal to  surfaces of constant $S$, from (\ref{4.10c}) we get an equation of motion for the curve:   
\begin{equation}
 (\partial_t+\frac{1}{m}\dot{\x}
 \cdot\boldsymbol{\nabla} ) m\dot{\x}=m\ddot{\x} = -\boldsymbol{\nabla} (V+Q).  \label{4.10e} 
\end{equation}  
This has the form of a classical equation of motion, but with the classical potential $V$ augmented by the quantity Q, commonly called the \textit{Quantum Potential} to emphasize its distinctive origin.
 
A striking fact about $Q$ is its influence on electron motion even in the absence of external forces. Its noteworthy use in \cite{Phillipidis79} to compute particle paths in electron diffraction  stimulated a resurgence of interest in Pilot Wave theory. That computation supported interpretation of $Q$ as a ''causal agent'' in diffraction, but identification of a plausible ''physical mechanism'' to explain it has remained elusive.  So interpretation of $Q$ as an intrinsic property of the wave function that does not require further explanation has remained the default position in Pilot Wave theory.
 
The Pauli equation has been used to analyze the effect of spin on electron paths in 2-slit diffraction 
\cite{Holland03}. The authors identify the correct generalization of the Pilot Wave \textit{guidance law} (\ref{4.10dd})
as
 \begin{equation}
 \dot{\x} =\boldsymbol{\nabla} S+\rho^{-1}\boldsymbol{\nabla}\boldsymbol{\times} (\rho \s)
   \label{4.10ee}  
 \end{equation}
However, they failed to note the 
more fundamental fact that, even in 
Schr\"odinger theory, the ``quantum force'' is spin dependent, though that was spelled out in one of their references
\cite{Hest71b}. Indeed that reference derived the equation of motion   
\begin{equation}
 \rho m\ddot{\x} = \rho\mathbf{f} +\grave{\mathbf{T}}(\grave{\boldsymbol{\nabla}}),
   \label{4.10f} 
\end{equation}  
where the accent indicates differentiation of the stress tensor $\mathbf{T}(\mathbf{n})$, and the applied force has the general form
\begin{equation}
\mathbf{f}=e[\E +\v \boldsymbol{\times}\B /c] +\frac{e}{mc}\grave{\boldsymbol{\nabla}}\grave{\B}\cdot \s, \label{4.10g} 
\end{equation}  
while components of the stress tensor are
\begin{equation}
\bsig_i\cdot\mathbf{T}(\bsig_j)
=\frac{\rho}{m}\s\bdot[\partial_i\partial_j \s+\s\,\partial_i\partial_j\ln \rho]=T_{ji}.
   \label{4.10h} 
\end{equation} 
When the spin vector $\s$ is constant, the stress tensor term in (\ref{4.10f}) reduces to the ``Quantum force'' $-\boldsymbol{\nabla} Q$ in Schr\"odinger theory. Thus we see that the $\hbar^2$ factor in Q comes from squaring the spin vector, and the Quantum force is actually a momentum flux. All this puts the diffraction problem in new light. Indeed, we shall see that spin dependence of the quantum force is even more obvious in Dirac theory.

Derivation of Pauli and Schr\"odinger equations as nonrelativistic approximations to the Dirac equation in \cite{Hest75b} also traces corresponding changes in local  observables. That brings to light many inconsistencies and omissions in standard treatments of those approximations.
The most egregious error is failure to recognize that the 
Schr\"odinger equation describes the electron in an eigenstate of spin. 
Implications of that fact are discussed at length in \cite{Hest79}.

Another surprising result from \cite{Hest75b,Hest79} is proof that $\beta$ makes an indisputable contribution to the energy in Pauli-Schr\"odinger theory, even though it has been banished from the wave function. It arises from the spin density divergence (\ref{B.17r}), which in the non-relativistic approximation
takes the form
\begin{equation}
m_{e}c\rho  \beta = - \boldsymbol{\nabla}\bdot (\rho \s).
 \label{4.10}
\end{equation}
This deepens the mystery of $\beta.$
More clues come from solutions to the Dirac equation.

\subsection{Pilot Waves in Dirac Theory}

Extension of the Pilot Wave interpretation for nonrelativistic wave functions \cite{Norsen13} to Dirac theory with STA has been critically examined at lenght in \cite{Doran96}, where it is demonstrated with many examples that calculations and analysis with the Real Dirac equation is no more complicated than with the Pauli equation. Indeed, the first order form of the Dirac equation makes some of it decidedly easier.
The treatment of scattering at potential steps is generalized to include both spin and oblique incidence, with STA simplifications not to be found elsewhere.
The analysis of evanescent waves exhibits the flow of Dirac streamlines (without commitment to their interpretation as particle paths).
The study of tunneling times
shows how part of the wave packet passes through the barrier while part slows down and turns back. No notion of wave function collapse is needed to interpret observations.
It is also shown that the distribution of tunneling times observed experimentally can be attributed entirely to structure of the initial wave packet, thus making it clear that, contrary to claims in the literature, no superluminal effects are involved.
The general conclusion is that interpretation of Dirac streamlines as particle paths is consistent with the Dirac equation and helpful in physical interpretation.

Indeed, the fundamental momentum balance equation (\ref{C.10}) gives us a complete and straightforward relativistic generalization of Pilot Wave theory that seems not to have been recognized heretofore. One needs only to apply it to a single streamline $z=z(\tau)$ with proper velocity $v=\dot{z}$ and spin bivector $S=S(z(\tau))$. Then the equation can be put in the form of a generalized \textit{Pilot Wave guidance equation}:
\begin{equation}
\square\Phi=m_{e}c\dot{z} + S\bdot \square \ln \rho+\dot{S}\bdot\dot{z},\label{4.11a}
\end{equation}
where 
\begin{equation}
\square\Phi = P-\frac{e }{c}A.\label{4.11b}
\end{equation}
is the gradient of a generalized electron phase expressed in action units. This gradient expression may have important implications for electron diffraction. For a free particle, the generalized momentum $P$ is necessarily a phase gradient. However, electron motion in diffraction  might also be influenced through a vector potential generated by material in the guiding slits. Since the curl $\square\wedge A$ must vanish in the vacuum near the slits, the vector potential is necessarily a gradient, so it can be combined with $P$ as in (\ref{4.11b}). This possibility has been overlooked in the literature on diffraction. It may be crucial for explaining how the slits transfer momentum to each electron in diffraction. We will have more to say about that in the Section on many particle theory.

The remaining piece of Pilot Wave theory
is given by the conservation law for the Dirac current as expressed by (\ref{C.11}). Evaluated on the particle path it gives us
\begin{equation}
\square^2\Phi=- m_e c\dot{z}\bdot \square \ln \rho,\label{4.11c}
\end{equation}
which describes the evolution of path density.

The relativistic guidance law (\ref{4.11a}) not only combines the
the two basic equations (\ref{4.10a}) and (\ref{4.10ee}) of nonrelativistic theory into one, it generalizes the scalar \textit{Quantum Potential} into a vector
$S\bdot \square \ln \rho$ and makes its spin dependence explicit. 

To compare the two versions, we must perform a spacetime split of (\ref{4.11a}), taking due account of their different notations. 
For brevity, we draw on the more complete treatment of spacetime splits in Section VIIB.
From (\ref{6.27}) we have the velocity split
\begin{equation}
\dot{z}\gamma_{0}=\gamma +\dot{\r}\qquad\hbox{with}
\qquad \gamma =c\dot{t}.\label{4.11d}
\end{equation}
and, for the spin bivector $S=isv$, from (\ref{6.33}) we have the  split
\begin{equation}
S=\s\boldsymbol{\times} \dot{\r}+i\s_{\perp} .\label{4.11e}
\end{equation}
where $\s_{\perp}$ is given by (\ref{6.33}).
Writing $a=\square \ln \rho$ with the split
$a\gamma_0=a_0+\a$ and using (\ref{2.23}), we get the split of the ``Quantum Vector Potential:'' 
\begin{equation}
(S\bdot a)\gamma_0 =
 (\s\boldsymbol{\times} \dot{\r})\bdot \a 
 +a_0
\s\boldsymbol{\times} \dot{\r}+\a\boldsymbol{\times} \s_{\perp} . \label{4.11f}
\end{equation}
Putting it all together, for the split of the guidance law (\ref{4.11a}), we get the generalization of (\ref{4.10a}) and (\ref{4.10ee}):
\begin{equation}
c^{-1}\partial_t\Phi=m_{e}c\gamma + (\s\boldsymbol{\times} \dot{\r})\bdot \boldsymbol{\nabla} \ln \rho ,\label{4.11ea}
\end{equation}
\begin{equation}
\boldsymbol{\nabla}\Phi=m_{e}c\,\dot{\r} + (c^{-1}\partial_t \ln \rho) \s\boldsymbol{\times} \dot{\r}-\s_{\perp}\boldsymbol{\times} \boldsymbol{\nabla}\ln \rho.\label{4.11eb}
\end{equation}
A detailed proof that the term $(\s\boldsymbol{\times} \dot{\r})\bdot \boldsymbol{\nabla} \ln \rho$ does indeed reduce to Bohm's quantum potential in the nonrelativistic limit is not needed here. Suffice it to say that both have been derived from Dirac's equation.
The term $\dot{S}\bdot\dot{z}$ has been ignored in these equations, because it has no analogue in the nonrelativistic theory. It's implications are studied in the following Sections.

\subsection{Stationary states with $\beta$}

To solve the Cauchy problem for an electron, it is convenient to perform a spacetime split of operators in the Dirac equation (\ref{4.1}) to put it in the form
\begin{equation}
(\partial_t +c\boldsymbol{\nabla}\Psi\i\hbar
=m_{e}c^2\Psi^* +e(A_0-\A)\Psi,\label{4.11}
\end{equation}
where $\Psi^*=\gamma_0\Psi \gamma_0$.
This equation is readily re-expressed in standard Hamiltonian form
\begin{equation}
\partial_t \Psi\i\hbar
=\underbar{H}\Psi,\label{4.12}
\end{equation}
though the structure of the Hamiltonian operator $\underbar{H}$ may look unfamiliar at first..

Boudet has applied this approach to a thorough treatment of the Darwin solutions for Hydrogen and their application to basic state transitions \cite{Boudet09}. (See also \cite{Doran96} for a somewhat different STA treatment.)
For a stationary state with constant energy $E$ and central potential $V(r)$, the wave function has the form
\begin{equation}
\Psi(\r,t)=\psi(\r) e^{-i\bsig_3Et/\hbar}.\label{4.13}
\end{equation}
And Boudet puts equation (\ref{4.11}) in the form
\begin{equation}
 \boldsymbol{\nabla}\psi = \frac{1}{\hbar c}[-E_0\psi^* +(E+V)\psi]i\bsig_3,  \label{4.14}
\end{equation}
where $E_0 = m_e c^2$.
He then splits the wave function into even and odd parts defined by 
\begin{equation}
\psi = \psi_e +i\psi_o \qquad \psi^* = \psi_e -i\psi_o
\label{4.15}
\end{equation}
to split (\ref{4.14}) into a pair of coupled equations for quaternionic spinors: 
\begin{equation}
 \boldsymbol{\nabla}\psi_e = \frac{1}{\hbar c}[-E_0 -E+V)\psi_o\bsig_3,  \label{4.16}
\end{equation}
\begin{equation}
 \boldsymbol{\nabla}\psi_o = \frac{1}{\hbar c}[-E_0 +E+V)\psi_e\bsig_3.  \label{4.17}
\end{equation}
These he solves to get the Darwin solutions. 

The same even-odd split was used in \cite{Hest75b} to get non-relativistic approximations to the Dirac equation. The split there mixes $\beta$ and boost factors in a peculiar way with no evident meaning.
Indeed, the peculiar behavior of  $\beta$ and local velocity in the Darwin solutions defies any sensible physical interpretation in terms of local observables,
with nodes separating positive and negative energy components in strange ways \cite{Gurtler72}.
These facts are not even recognized in the standard literature, let alone regarded as problematic.
Nevertheless, they pose a challenge to associating particle properties with the wave function.

To address this challenge, we first look to clarify the interpretation of $\beta$ in the simpler case of a free particle wave packet. Then we consider a new approach to relating particle properties to the wave function in the next Section.

We look for wave packet solutions of the free particle wave equation
\begin{equation}
(\partial_t +c\boldsymbol{\nabla})\Psi i\bsig_3\hbar
=m_{e}c^2\Psi^* \label{4.18}
\end{equation}
that generalize the 
plane wave solutions with zitter in Section IVC.
Accordingly, we look for solutions of the form
\begin{equation}
\Psi(t,\r) =(\rho e^{i\beta})^{1/2} R = e^{\half(\alpha +i\beta)-i\bsig_3 \varphi}\, V_0\, e^{-i\bsig_3 \theta} \label{4.19},
\end{equation}
where $V_0 = a + b \bsig_2$ is a constant boost and
$\varphi=p\bdot x/\hbar$ with $p=m_e c\gamma_0$.
For the simplest case with $V_0=1$, substitution into the wave equation gives us the complex equation of constraint: 
\begin{align}
\dot{\alpha} +i\dot{\beta}+\boldsymbol{\nabla} \alpha +&i\boldsymbol{\nabla} \beta
-2(\dot{\theta}+\boldsymbol{\nabla} \theta)i\bsig_3 \notag \\
 & =\lambda_e^{-1} [1-e^{-i\beta}]i\bsig_3,\label{4.20}
\end{align}
where $\lambda_e =\hbar/2m_e c$. Separating out scalar and bivector parts, we get the equations
\begin{equation}
\dot{\alpha} =0,\label{4.21}
\end{equation}
\begin{equation}
\boldsymbol{\nabla} \beta
-2\,\dot{\theta}\bsig_3 
  =\lambda_e^{-1} [1-\cos\beta]\bsig_3,\label{4.22}
\end{equation}   
while vector and pseudoscalar parts give us
\begin{equation}
\boldsymbol{\nabla} \alpha =
2\,\bsig_3\boldsymbol{\times} \boldsymbol{\nabla} \theta 
  +\lambda_e^{-1} \bsig_3 \sin\beta,\label{4.23}
\end{equation}
\begin{equation}
\dot{\beta}
=2\bsig_3 \bdot \boldsymbol{\nabla} \theta . \label{4.24}
\end{equation}
These four constraints on components of the wave function 
are part of the constraints on any solution of the Dirac equation, so what they tell us is of general interest.

First, equation (\ref{4.21}) tells us that the density $\rho$ is constant in time, so $\alpha =\ln \rho = \alpha(\r)$ is a function of position only. Its shape is determined by initial conditions consistent with the other constraints. But wait, it is a theorem that any Dirac wave packet must expand in time. Indeed, no  closed wave packet solutions of the free particle Dirac equation are known \cite{Park12}. 
However, conventional wave packets are superpositions of plane wave solutions with momenta in different directions causing the spread. And in this case the momentum is the same at every point of the packet at a given time, which seems to escape the spreading theorem.

Now here is the informative part. Inspection of eqns. (\ref{4.22}) and (\ref{4.23}) shows us that  $ \beta$ and $\square \theta$ determine independent components of $\square \alpha$. More to the point, for a given a density function with the phase gradient $\square \theta$ balancing $\square \alpha$ in the spin plane, the function $\beta$ can be always adjusted to balance the component of $\square \alpha$ in the spin direction. So to speak, it appears that the role of $\beta$ is to adjust the probability density to the spin. 
Thus, we have considerable freedom in adjusting the functional form of the probability density to given initial conditions, such as spherical symmetry for example.

This fact raises a question about  the $\beta$-factor in Born-Dirac theory. Should it be lumped with probability density $\rho$, so $(\rho e^{i\beta})^{1/2}$ is the full probability component of the wave function? Or should it be lumped with the rotor $R$ to describe the kinematics of motion? A surprising answer is given below. However, it does not preclude the possibility that $\beta$ might also play the role of chirality parameter.

\subsection{Scattering and QED with zitter}

The link between standard quantum mechanics (QM) and quantum electrodynamics (QED) passes through the Dirac equation. It is commonly claimed that the link requires second quantization with quantum field theory (QFT). But Feynman vehemently denied that claim. When the issue arose in a QED course I attended, I recall him dramatically remonstrating that, if anyone dares to defend axioms of QFT, ``I will defeat him. I will CUT HIS FEET OFF!'' (with a violent cutting gesture for emphasis).  
Indeed, the famous formula $[p,q] = i\hbar$, which Born proposed as a foundation of QM (and had engraved on his tombstone), cannot be as general as he thought. For there is no explanation why Planck's constant here is related to electron spin or the Dirac equation. 
Also, one can argue that QFT commutation relations for particle creation and annihilation operators are merely bookkeeping devices for multiparticle physics without introducing new physics.
Let’s look at how Feynman got along without it.

A reformulation of Feynman's approach to QED with STA is laid out in \cite{Lewis2000,Lewis2001}, with explicit demonstrations of its advantages in Coulomb and Compton scattering calculations.
For example, the $S$-matrix is replaced by a scattering operator $S_{fi}$ that rotates and dilates the initial state to the final state, as expressed by
\begin{equation}
\psi_f=S_{fi} \psi_i \label{4.25}
\end{equation}
with  
\begin{equation}
S_{fi}=\rho_{fi}^{1/2} R_{fi}, \label{4.26}
\end{equation}
where $R_{fi}$ is a rotor determining the change in direction of spin as well as momentum, while $\rho_{fi}=|S_{fi}|^2$ is a scalar dilation factor determining the cross section.

Feynman linked QM to QED by reformulating the 
Dirac equation as an integral equation 
coupled to Maxwell theory through the vector potential:
\begin{equation}
\psi(x)=\psi_i(x)-e\int d^4 x'S_F(x-x')A(x')\psi(x'). \label{4.27}
\end{equation}
This solves the Dirac equation (\ref{4.3}) with $p_0=m_e c\gamma_0$ if the Green's function $S_F(x-x')$ satisfies the equation
\begin{align}
\square S_F(x-x')M(x')\i & - S_F(x-x')M(x')p_0 \notag \\ 
& =\varphi^4(x-x')M(x'),\label{4.28}
\end{align}
where $M=M(x)$ is an arbitrary multivector valued function of $x$. It has the \textit{causal solution}
\begin{align}
 S_F(x-x')M\i  =&- \frac{\Theta(t-t')}{(2\pi)^3}\int\frac{d^3 p}
{2E }(pM +Mp_0)\i e^{-\i p\cdot (x-x')} \notag \\ 
& +\frac{\Theta(t-t')}{(2\pi)^3}\int\frac{d^3 p}{2E }(pM +Mp_0)\i e^{\i p\cdot(x-x')},
\label{4.29}
\end{align}
where $E= p\cdot\gamma_0 >0$. Note that $S_F(x-x')$ is a linear operator on $M$ here. In general $M$ does not commute with $p$,  $p_0$, or the bivector $\i=i\bsig_3$, so it cannot be pulled from under the integral.

We can draw several important conclusions from the present approach to QED. One advantage of the integral form (\ref{4.27}) for the Dirac equation is that the causal boundary condition  (\ref{4.29}) explicitly enforces the association of electron/positron states with positive/negative energy states respectively. As noted in Section IVB, these states can be switched by multiplication with the pseudoscalar $i$.

At this point, permit me to insert a relevant anecdote that I heard Feynman tell on himself.
One day, when he was demonstrating his spectacular prowess at complex QED calculations, a brave student objected: ``You can't normalize negative energy states to plus one, you must use negative one.'' ``O yes I can!'' retorted Feynman with the confidence of one who had won a Nobel prize with his calculations and demonstrated them repeatedly over more than a decade in QED courses and lectures. Then he proceeded to prove that the student was right! Sure enough, check out eqn. (\ref{A.8}) to see that the minus sign comes from squaring the unit pseudoscalar (which, of course,  Feynman never did learn)!

Returning to the main point, we note that the absence of 
a $\beta$-factor $e^{i\beta}$ in the scattering operator (\ref{4.26}) shows that positive and negative energy  states are not mixed in scattering.
Indeed, the question of a $\beta$-factor never arises in QED, because all calculations are based on plane waves without it, and it is not generated by conventional wave packet construction.

Of course, the Born rule is not an intrinsic feature of the Dirac equation, but is imposed only for purposes of interpretation.
It is important, therefore, to recognize that results of plane wave scattering have a straight forward geometric interpretation without appeal to probability:
Indeed, the Dirac equation generates a unique spacetime path for each point on an initial plane wave. The conservation law for the Dirac current implies that these paths do not intersect, though they may converge or separate.
Accordingly, if we assign uniform density to paths beginning on the initial plane wave, then the scattering operator determines the density of particle paths intersecting a surface surrounding the scattering center.
In other words. \textit{the squared modulus $\rho$ of the Dirac wave function specifies the density of particle paths!}
This is a completely geometric result, independent of any association with  probabilities. 
Of course, for experimental purposes the density of paths can be interpreted as a particle probability density, but no inherent randomness in nature is thereby implied.

The bottom line is that QED scattering is fundamentally about paths.

Our STA formulation reveals another aspect of QED that has been generally overlooked and may be fundamental; namely, the existence of zitter solutions and the possibility that they may describe a fundamental feature of the electron.
As we have seen, zitter wave functions with opposite chirality can be obtained from a general wave function $\Psi$ by projection with a lightlike
 ``zitter boost''
\begin{equation}
\Sigma_\pm =\gamma_\pm\gamma_{0}= (\gamma_{0}\pm \gamma_2)\gamma_{0}= 1 \pm \bsig_2. \label{4.31}
\end{equation}
Thus we obtain
\begin{equation}
\Psi_\pm (x)= \Psi (x)\Sigma_\pm =
(\rho e^{i\beta})^{1/2}R\,\Sigma_\pm, \label{4.32}
\end{equation}
where, as before, $R=R(x)$ is a general spacetime rotor, though we often wish to make the phase explicit by writing $R=Ve^{-\i\varphi}$. 
Actually, the $\beta$-factor can also be incorporated into the rotor $R$, to give us
\begin{equation}
e^{i\beta/2}Ve^{\i\varphi}\Sigma_+=Ve^{i\bsig_3\varphi}e^{i\bsig_2\beta/2}\Sigma_+ , \label{4.33}
\end{equation}
 because the $\Sigma_+$ factor converts it to a rotation:   
\begin{equation}
e^{i\beta/2}\Sigma_+=e^{i\bsig_2\beta/2}\Sigma_+ =\Sigma_+e^{i\bsig_2\beta/2}. \label{4.34}
\end{equation}
Note that the $\beta$-rotation will occur before the phase-rotation in expressions for local observables given below. Thus, \textit{the $\beta$-factor tilts the spin vector before the phase rotation in the spin plane.}  In other words, it is a ``tilting factor.''
Here, at last, we have a clear geometric meaning for the parameter $\beta$! It suggests that the ``$\beta$ problem'' for Hydrogen can be solved by simply multiplying Boudet's Darwin solutions by a zitter boost to get zitter solutions. The physical significance of zitter in QED remains an open question. 
But later we will note a possible role in  weak interactions. 

Zitter boosts possess the reversion, idempotence and orthogonality properties
\begin{equation}
\widetilde{\Sigma}_\pm =\Sigma_\mp,\quad
(\Sigma_\pm)^2 =2\Sigma_\pm,\quad
\Sigma_\pm\widetilde{\Sigma}_\pm  =0. \label{4.35}
\end{equation}
Consequently, we have lightlike local observables for electron current :
\begin{equation}
\half \Psi_+\gamma_0 \widetilde{\Psi}_+ 
=\Psi(\gamma_0+\gamma_2 )\widetilde{\Psi} =\rho u,
 \label{4.36}
\end{equation}
and for spin:
\begin{equation}
\half \Psi_+i\bsig_3 \widetilde{\Psi}_+=\Psi(\gamma_0+\gamma_2 )\gamma_1\widetilde{\Psi}=\rho ue_1,
 \label{4.37}
\end{equation}
or
\begin{equation}
\rho S=\hbar\Psi_{+}i\bsig_{3}\widetilde\Psi_{+}=\half\hbar\Psi\gamma_{3}\gamma_{+}\widetilde\Psi=\rho su.  \label{4.37a}
\end{equation}

Now, to adapt the Dirac equation (\ref{4.3}) to zitter and relate it to local observables, we try projecting it into the form
\begin{equation}
(\hbar\square\Psi_+ i\bsig_3
-\frac{e }{c}A\Psi_+)\widetilde{\Psi}_+
=m_{e}c\Psi_+\gamma_0\widetilde{\Psi}_+.\label{4.38}
\end{equation}
This fails, however, because 
$ \Psi_+\widetilde{\Psi}_+=0$. One way to correct that is to modify the left side to the gauge invariant form
\begin{equation}
(\hbar\square
+\frac{e }{c}AI)\Psi_+ i\bsig_3\widetilde{\Psi}_+,
\label{4.39}
\end{equation}
with $I=Vi\bsig_3\widetilde{V} $, so
\begin{equation}
-I\Psi_+ i\bsig_3\widetilde{\Psi}_+=\Psi_-\widetilde{\Psi}_+ =2\rho (1-e_2 e_0).
\label{4.40}
\end{equation}
With a similar adaptation of (\ref{B.39r}),
we get
\begin{equation}
\frac{\hbar}{2} (\square\Psi_+ )i\bsig_3\widetilde{\Psi}_+
=\rho P(1-e_2 e_0) +\square(\rho S).\label{4.41}
\end{equation}
Putting it all together in analogy to 
(\ref{B.40r}), we get the equivalent of the zitter Dirac equation in terms of local observables: 
\begin{equation}
\rho (P-\frac{e }{c}A )(1-e_2 e_0)=m_{e}c\rho u
-\square(\rho S),\label{4.42}
\end{equation}
This can be separated into a trivector part
\begin{equation}
\rho (P-\frac{e }{c}A )\wedge (e_2 e_0)
=\square\wedge(\rho S),\label{4.43}
\end{equation}
and a vector part
\begin{equation}
\rho (P-\frac{e }{c}A )-\rho 
(P-\frac{e }{c}A )\bdot (e_2 e_0)
=m_{e}c\rho u
-\square\bdot (\rho S),\label{4.44}
\end{equation}
wherein we recognize the zitter version of the Gordon current.

Averaging over a zitter period reduces (\ref{4.42}) to the Gordon current in (\ref{C.5}), which we have already identified as of prime physical importance. And that has the clear physical meaning of averaging out zitter fluctuations, in contrast to simply dropping the $\beta$-factor as done before.

Furthermore, the trivector equation (\ref{4.43}), like its zitter average (\ref{C.6}), puts a constraint on the observables with no evident physical significance. So we have good reason to drop it altogether. Fortunately, that can done in a principled way, simply by symmetrizing equation (\ref{4.42}) with its reverse, which eliminates the trivector part. This is tantamount to declaring the Gordon current (with zitter) as physically more significant than the Dirac equation. 

The bottom line is a claim that observables of the wave function $\Psi_+(x)$ describe a congruence (or fibration, if you will) of lightlike helical paths with the circular period of an electron clock.
Then we aim to extract individual fibrations from the wave function to create a well-defined particle model of electron motion.

\section{ Pilot Particle Model}\label{sec:}

We assume that the lightlike helical path of a fiber in the wave function has a well-defined center of curvature with a timelike path  that we can identify as a particle Center of Mass (CM).
Accordingly, we regard the CM  as a \textit{particle}
with intrinsic spin and internal clock.
Then we can describe it by observables and equations of motion derived from the Dirac theory in the preceding Section.
With deference to de Broglie, let's call the resulting particle model a \textit{Pilot Particle} embedded in a \textit{Pilot Wave} solution of the Dirac equation.

Our first task in this Section is to derive and study equations of motion for a well-defined and self-consistent  \textit{Pilot Particle Model} (PPM).
Our second task is to apply the PPM to  description of quantized  stationary states.
Remarkably, our derived quantization conditions are identical to those of Sommerfeld  using Old Quantum Theory. However, inclusion of spin in our particle model enables resolution of famous discrepancies in Sommerfeld's energy spectrum for Hydrogen when compared with the standard result derived from the Dirac equation.

\subsection{Equations for Particle Motion with Clock}

Let $ z=z(\tau) $ designate the \textit{particle path} with  \textit{proper velocity} $ v=\dot{z} =e_{0} $.  
We have already identified this path with the CM of an electron with zitter and derived an exact equation for it 
(\ref{C.10}). That equation is so important, it bears repeating for comparison with the approach in this Section:
\begin{equation}
 p=m_{e}c v+ \dot{S}\bdot v=P-\frac{e}{c}A.\label{6.01}
\end{equation}
Here we begin anew to derive and analyze stand-alone particle equations of motion from
the treatment of local observables in the preceding Section.

The electron's local observables  are now restricted to a comoving frame attached to the path:
\begin{equation}
e_{\mu}=e_{\mu}(\tau)=R\gamma_{\mu}\tR,\label{6.2}
\end{equation}
where $ R=R(\tau) $ is a Rotor with spin vector $ s=(\hbar/2)e_{3}$ and spin bivector $S=isv$ defined as before.
 
Our problem  is to ascertain dynamical consequences of the Dirac equation for the Pilot Particle. 
We seek equations of motion consistent with the conservation laws of Dirac theory
formulated in terms of local observables, for the most part under the assumption $ \cos\beta=1 $.
As our analysis is not a straightforward derivation, our results may not be completely general, but we take care to make them self consistent.

On the particle path, the particle momenta $p$ and $p_{c}$  retain their forms in (\ref{B.38r})  and (\ref{B.38rr}),
so we can simply repeat them here for reference:
\begin{equation}
p=P-\frac{e}{c}A,\label{6.5}
\end{equation}
 \begin{equation}
p_{c}=(p\bdot v) v+\dot{S}\bdot v - s\dot{\beta},\label{6.5a}
\end{equation}  
where 
\begin{equation}
p\bdot v
=\Omega\bdot S=p_{c}\bdot v.\label{6.5g}
\end{equation}
can be regarded as a (possibly variable) \textit{dynamical mass}.
The problem remains to relate $\Omega$ to particle dynamics.
Also, it should be understood that the ``\textit{spin momentum}'' term $ \dot{S}\bdot v=\dot{v}\bdot S$ describes linear momentum due to internal angular momentum, like a flywheel in a macroscopic moving body.
The term $ s\dot{\beta}$ has been retained for generality, though we shall mostly ignore it in the following.

Ascertaining equations for momentum and angular momentum conservation on the particle path is simplified by  the assumption that particle density is constant on the path, so we  can  factor it out of the exact equations.
One way to derive conservation laws for the particle motion is to surround the particle path with a spherical tube. Then, integrating 
the equation for momentum conservation (\ref{B.21r}) over the tube and, assuming no radiation through the spherical walls, shrinking the tube to the particle position $ z(\tau ) $ gives the familiar Lorentz force equation
\begin{equation}
\dot{p}=\dot{p}_{c}= \frac{e }{c}F\bdot v,\label{6.6}
\end{equation}
where $F= F(z(\tau)) $ is the external field acting on the particle.

Similarly, equations (\ref{B.23ar}) and (\ref{B.24br}) for angular momentum conservation give us  
\begin{equation}
\dot{J}=\dot{p}\wedge z \qquad \hbox{and}\qquad  \dot{J}_{c}=\dot{p}_{c}\wedge z,\label{6.7}
\end{equation}
where, in accordance with (\ref{B.23cr}) and (\ref{B.24cr}), the total angular momentum is defined by 
\begin{equation}
J=p\wedge z+S\qquad \hbox{or}\qquad J_{c}=p_{c}\wedge z-S.\label{6.8}
\end{equation}
Inserting (\ref{6.8}) into (\ref{6.7}) yields  alternative equations for angular momentum conservation:
\begin{equation}
\dot{S}=p\wedge v=v \wedge p_{c}.\label{6.9}
\end{equation}
An explicit relation between the two kinds of momenta $p$ and $p_{c}$ can be derived by dotting this double equation with $v$ and using (\ref{6.5a}); whence 
\begin{equation}
p=p_{c}-2\dot{S}\bdot v=(p\bdot v) v-
\dot{S}\bdot v .\label{6.9m}
\end{equation}

Note that equation (\ref{6.9}) implies the constraint
\begin{equation}
\dot{S}\wedge v=0.\label{6.9n}
\end{equation}
We can use $ S=isv $ to put it in the form
\begin{equation}
(\dot{s}\wedge v +s\wedge \dot{v})\bdot v=(\dot{s}\wedge v)\bdot v=0.\label{6.9a}
\end{equation}
Hence,
\begin{equation}
\dot{s}=(\dot{s}\bdot v)v=-(s\bdot \dot{v})v.\label{6.9b}
\end{equation}
Or equivalently,
\begin{equation}
\dot{S}=i(s\wedge \dot{v}).\label{6.9c}
\end{equation}
Thus, the spin bivector is constant if and only if the particle acceleration is collinear with the spin vector.

Instead of appealing to conservation laws for particle equations of motion, we can look directly to the Dirac equation.  Analysis is simplified by assuming that for any function $f=f(z(\tau))$ defined on the particle path,
\begin{equation}
v\bdot \square f=\dot{f}\quad\hbox{and}\quad \square f=v \dot{f}.\label{6.9ca}
\end{equation}
In particular, for momentum on the particle path, from (\ref{C.45}) we get
\begin{equation}
v\wedge\dot{p} =-\frac{e }{c} F+\square\wedge P,\label{6.9cb}
\end{equation}
Hence,
\begin{equation}
\dot{p} =\frac{e }{c} F\bdot v+v(v\bdot \dot{p})-v\bdot (\square\wedge P)\label{6.9cbb}
\end{equation}
This agrees with our force law (\ref{6.6}) only if 
\begin{equation}
v\bdot (\square\wedge P)=0.\label{6.9cc}
\end{equation}
Perhaps this is a more general condition for radiationless motion than $\square\wedge P=0$, which we have already associated with stationary states. Recall that standard interpretations of Dirac theory exclude radiation, regarding it as a special province of QED. No such exclusion is assumed here, though questions about radiation are deferred for another time.

Now note that we can solve (\ref{6.9}) for
\begin{equation}
p=(p\bdot v+p\wedge v) v=(p\bdot v)v+\dot{S}\bdot v\label{6.6ff}
\end{equation}
And if we apply (\ref{6.9ca}) to restrict
the Gordon current (\ref{C.10}) to the particle path we get:
\begin{equation}
 p=P-\frac{e }{c}A
 =m_{e}c v-v\bdot\dot{S}.\label{6.6g}
\end{equation}
Thus the Gordon current tells us that the dynamical mass $p\bdot v$ can be reduced to the rest mass, and $p$ (rather than $p_{c}$)  should be regarded as the particle momentum.

While the \textit{kinetic momentum} $m_{e}c v$ in (\ref{6.6g}) is familiar the \textit{spin momentum} $q\equiv \dot{S}\bdot v=-v\bdot\dot{S}$ is not, so let us examine some of its properties. 
Using (\ref{6.9c}) we can put it into the form 
\begin{equation}
q=\dot{S}\bdot v=[i(s\wedge \dot{v})]\bdot v=i(s\wedge\dot{v}\wedge v)=(i\dot{v}v)\bdot s.\label{6.6v}
\end{equation}
Note that we can drop the wedge in 
$s\wedge\dot{v}\wedge v=(s\wedge\dot{v}) v$ because $v$ is orthogonal to the other two vectors. Hence,
\begin{equation}
q^2=(\dot{S}\bdot v)^2
= -(s\wedge\dot{v})^2=s^2\dot{v}^2 -(s\bdot\dot{v})^2.\label{6.6w}
\end{equation}
Spin and kinetic momenta are orthogonal to one another, because
\begin{equation}
q\bdot v=(\dot{S}\bdot v)\bdot v
=\dot{S}\bdot (v\wedge v)=0.\label{6.6vv}
\end{equation}
Hence,
\begin{equation}
p^2=(m_e cv)^2+(\dot{S}\bdot v)^2= m^2_e c^2-(s\wedge\dot{v})^2.\label{6.6x}
\end{equation}
This suggests that
\begin{equation}
-(s\wedge\dot{v})^2=(\dot{S}\bdot v)^2=-(p\wedge v)^2\label{6.6y}
\end{equation}
is a measure of energy (or mass) stored ``in'' an accelerated electron, so its time derivative is a measure of radiated energy. 
From the Lorentz force we find an equation for its flow:
\begin{equation}
\dot{p}\bdot p=\frac{-1}{2c}\frac{d}{d\tau}(p\wedge v)^2= \frac{e }{c}F\bdot(v\wedge p)\label{6.6z}
\end{equation}
For acceleration of spin momentum on its own, we find from (\ref{6.6v}):
\begin{equation}
\dot{q}=i[(s\wedge\dot{v})^{\bdot}\wedge v]\quad \hbox{so}\quad \dot{q}\bdot v=0.
\label{6.6va}
\end{equation}
But questions about the physical significance of spin momentum will remain until testible physical implications are worked out and verified.

Turning now to a different issue,
kinematics of the comoving particle frame (\ref{6.2})
is determined by an angular velocity bivector 
\begin{equation}
\Omega\equiv \Omega(z(\tau)) =2\dot{R}\tR,\label{6.3}
\end{equation}
so
\begin{equation}
\dot{e}_{\mu}=\Omega\bdot e_{\mu}\label{6.4}
\end{equation}
on the electron path.
Using $\square\wedge v =v\dot{v}$ in (\ref{C.33}) we get  
\begin{equation}
\Omega(z(\tau))=\dot{v} v
+(P\bdot v) S^{-1},\label{6.6l}
\end{equation}
Hence, for motion of the ``electron clock hand'' $e_{1}$ we have
\begin{align}
\dot{e}_{1} &=(\dot{v}v)\bdot e_{1}+(P\bdot v) S^{-1}\bdot {e}_{1}\notag\\
&=-(\dot{v}\bdot e_{1})v
 +\frac{2P\bdot v}{\hbar}e_{2}. \label{6.6u}
\end{align}
This shows the role of the canonical momentum $P$ in evolution of the electron clock explicitly.
We aver that this mechanism determines quantization of atomic states, though we will apply it indirectly.

We have seen good reasons throughout this paper to regard the \textit{Canonical Momentum} $P$ in Dirac theory  as the \textit{Total Momentum} of the electron plus external sources, or, if you will, $\rho P$ as \textit{momentum density of the vacuum}. Accordingly, for
\begin{equation}
P=p+\frac{e}{c}A=m_{e}c v+\frac{e}{c}A+\dot{S}\bdot v,\label{6.6r}
\end{equation}
from (\ref{6.6}) we get the vacuum conservation law on the electron path
\begin{equation}
\dot{P}= \frac{e}{c}\,\grave{\square}\grave{A}\bdot v,\label{6.6q}
\end{equation}
where the emphasis on the right indicates which quantity is to be differentiated, and
 $v\bdot(\square\wedge A) =\dot{A}-\grave{\square}\grave{A}\bdot v$
has been used.
The external potential is said to be static in a reference frame specified by a constant vector $\gamma_{0}$ if $\gamma_{0}\bdot\square A=c^{-1}\partial_{t}A=0.$ In that case, we have a conserved total energy given by 
\begin{equation}
cP_{0}=cP\bdot\gamma_{0}=m_{e}c^{2}v\bdot\gamma_{0} +\frac{e}{c}A_{0}
+(\dot{S}\bdot v)\bdot\gamma_{0}.\label{6.6p}
\end{equation}
With some analysis, the last term can be identified as spin-orbit energy.

As developed to this point, our Pilot Particle model has much in common with  classical models for a ``particle with spin'' considered by many authors \cite{Hest10}, so it is of interest to see what they can contribute to our analysis. It is reassuring to know that the self consistency of those models was established by derivation from a Lagrangian. Since the kinematic details align perfectly with our present model, we can restrict our attention to the key kinematical equation studied there. 
 
The relevant rotor equation of motion in \cite{Hest10} has the strange but simple form:
\begin{equation}
\hbar\dot{R}\gamma_{2}\gamma_{1}= pR\gamma_{0}+\beta iR,\label{6.6a}
\end{equation}
where a (suggestive) dummy parameter $\beta$ has been introduced as placeholder for a simple identity derived below.
Using  (\ref{6.3}) we find an informative expression relating local observables:
\begin{equation}
\Omega S=pv+i\beta.\label{6.6b}
\end{equation}
We get several relations by decomposing it into pseudoscalar, bivector and scalar parts:
\begin{equation}
\Omega\wedge S=i\,\Omega\bdot(s\wedge v)=i(\dot{s}\bdot v),\label{6.6c}
\end{equation}
\begin{equation}
\dot{S}=\Omega\boldsymbol{\times} S= p\wedge v,\label{6.6d}
\end{equation}
\begin{equation}
\Omega\bdot S=p\bdot v.\label{6.6e}
\end{equation}
The pseudoscalar part is a mere identity.
However, the bivector part reproduces the spinor equation of motion (\ref{6.9}), while the scalar part 
gives us the same expression for  \textit{dynamical mass} as (\ref{6.5g}).

Of course, the dynamical mass $p\bdot v=\Omega\bdot S$ must be positive, which fixes the relative orientation of the two bivectors $\Omega$ and $S$. Suppose that is also true in a corresponding equation for a positron:
\begin{equation}
p\bdot v=-\Omega\bdot S,\label{6.6zz}
\end{equation} 
The opposite sign on the right indicates an opposite sense for rotational velocity $\Omega$ projected on the spin plane, in other words, the opposite chirality that distinguishes positron from electron. 

This completes our formulation of Pilot Particle  dynamics and kinematics. Its significance will be tested by application to a central problem of quantum mechanics, the description of quantized electron states. 
Physical interpretation of the results is facilitated by employing spacetime splits, which we turn to next.

\subsection{Particle Spacetime Splits}

Spacetime splits with respect to a timelike unit vector $ \gamma_{0} $ were defined in Section II. Here we are concerned with spacetime splits of particle equations where $ \gamma_{0} $ defines an inertial reference frame, called the ``Lab frame" for convenience. These splits are important for two reasons. First,  physical interpretation is enhanced, because of our experience with physics in inertial systems. Second, as we shall see, because quantization of stationary states is defined with respect to an inertial system. Here, we lay out the splits of relevant physical quantities and equations for convenient reference.

With the split of an arbitrary spacetime point $ x $ defined by 
(\ref{2.14}), the split of the particle path $ z=z(\tau) $ with respect to any arbitrary fixed origin $ x_{0} $ is given by 
\begin{equation}
(z-x_{0})\gamma_{0}=ct+\r,\label{6.25}
\end{equation}
where $ct=(z-x_{0})\bdot\gamma_{0}$ and
\begin{equation}
\r=(z-x_{0})\wedge \gamma_{0}=\z-\x_{0}.\label{6.26}
\end{equation}
Without loss of generality, we set $x_{0}=0$ hereafter.
Split of the proper velocity $ v=\dot{z} $ is then given by
\begin{equation}
v\gamma_{0}=c\dot{t}+\dot{\r}=\gamma(1+\v/c),\label{6.27}
\end{equation}
where
\begin{equation}
\dot{\r}=\frac{\gamma}{c}\v,\qquad \v=\frac{d\r}{dt},\qquad \gamma=c\dot{t}=\frac{dt}{d\tau}.\label{6.28}
\end{equation}
These splits define the variables for particle kinematics.
As they suggest, it is often simpler to use $ \dot{\r} $ rather than $ \v $ as velocity variable.

Splits of the spin vector and bivector are especially tricky. 
The split
\begin{equation}
s\gamma_{0}=s_{0}+\s\label{6.29}
\end{equation}
is simple enough. But 
\begin{equation}
s\bdot v=0=\gamma s_{0}-
\dot{\r}\bdot \s\label{6.30}
\end{equation}
implies
\begin{equation}
 s_{0}=s\bdot \gamma_{0 }=\frac{\s\bdot\v}{c}=
 \frac{\s\bdot\dot{\r}}{\gamma} .\label{6.31}
\end{equation}
Hence,
\begin{equation}
sv=s\wedge v=(s_{0}+\s)(\gamma-\dot{\r})=\s_{\perp}+\dot{\r}\wedge \s,\label{6.32}
\end{equation}
where
\begin{equation}
\s_{\perp}=\gamma\s-s_{0}\dot{\r}.\label{6.33a}
\end{equation}
So the split of the spin bivector $ S=isv $ is given by
\begin{equation}
S=\s\boldsymbol{\times} \dot{\r}+i\s_{\perp} .\label{6.33}
\end{equation}
Now 
\begin{equation}
S^{2}=s^{2}=\frac{-\hbar^{2}}{4}=s_{0}^{2}-\s^{2}=
(\s\boldsymbol{\times} \dot{\r})^{2}-\s_{\perp}^{2}.\label{6.34}
\end{equation}
Hence,
\begin{equation}
|\s|=[s_{0}^{2}-s^{2}]^{\half }=\frac{\hbar}{2} \left[ 1+\left( \frac{\hat{\s}\bdot \v}{c}\right) ^{2}\right] ^{\half}  .\label{6.35}
\end{equation}
Thus we have $ |\s|=|s|=\hbar/2 $ when ``helicity" $ \s\bdot\hat{\v}=0. $ Then the spin is said to be ``transverse" and $\s_{\perp}=\s.  $ When $ \s\boldsymbol{\times}\v=0, $ we have $ \s\bdot\hat{\v} =\pm|\s|,$ and the spin is said to be ``longitudinal." We shall see that spin is transverse in quantized states.

The split of the proper momentum  is given by 
\begin{equation}
p\gamma_{0}=p_{0}+\p\qquad \hbox{where}\qquad 
\p=p\wedge \gamma_{0}.\label{6.36}
\end{equation}
Whence the orbital angular momentum split is
\begin{align}
p\wedge z&=p_{0}\r-ct\p+i\mathbf{l}\notag\\ 
\hbox{with}\quad\qquad\qquad 
i\mathbf{l}&=\r\wedge\p=i(\r\boldsymbol{\times}\p).\label{6.37}
\end{align}
So the split of the total angular momentum 
$J=p\wedge z+ S$ can be put in the form  
$ J=\mathbf{j}_{0}+ i\mathbf{j} $, where  
\begin{equation}
\mathbf{j}_{0}=p_{0}\r -ct\p + \s\boldsymbol{\times}\dot{\r}\label{6.38}
\end{equation}
and 
\begin{equation}
\mathbf{j}=\mathbf{l} + \s_{\perp}=\r\boldsymbol{\times}\p  + \s_{\perp}.\label{6.40}
\end{equation}
Now we turn to splits for equations of motion.

Given the familiar split $ F=\E+i\B $, from  (\ref{2.23}) we get the split for the Lorentz force:
\begin{equation}
(F\bdot v)\gamma_0 = \E\bdot \dot{\r} +\gamma\E
+\dot{\r}\boldsymbol{\times} \B . \label{6.41}
\end{equation}
Accordingly, split of the momentum equation gives us
\begin{equation}
\dot{p}_{0}=\frac{e}{c}\E\bdot \dot{\r}\quad \hbox{with}\quad 
\dot{\p}=\frac{e}{c}(\gamma\E
+\dot{\r}\boldsymbol{\times} \B ).\label{6.42}
\end{equation}
The most helpful angular momentum equation is 
\begin{equation}
\dot{\mathbf{j}}=\dot{\mathbf{l}}+\dot{\s}_{\perp}=\frac{e}{c}
\r\boldsymbol{\times}(\E+\frac{\v}{c}\boldsymbol{\times} \B ) .
\label{6.44}
\end{equation}
Alternatively, with
\begin{equation}
vp=(\gamma+\dot{\r})(p_{0}-\p)
=p_{0}\dot{\r}
-\gamma \p+i(\dot{\r}\boldsymbol{\times} \p)+m_{e}c\notag\\
\end{equation}
and the split (\ref{6.33}), we get
\begin{equation}
(\s\boldsymbol{\times} \dot{\r})^{\textbf{.}}=p_{0}\dot{\r}
-\gamma \p .\label{6.46}
\end{equation}
and
\begin{equation}
\dot{\s}_{\perp}=\dot{\r}\boldsymbol{\times} \p.\label{6.47}
\end{equation}
Finally, we will be interested in the split
\begin{equation}
q\gamma_0=(\dot{S}\bdot v)\gamma_0
 = (\ddot{\r}\boldsymbol{\times} \dot{\r})\bdot\s 
 +\dot{\gamma}\,\s\boldsymbol{\times} \dot{\r}
+\gamma\,\ddot{\r}\boldsymbol{\times} \s_{\perp} . \label{6.47a}
\end{equation}
The scalar part is especially interesting, because it contributes a ``spin-orbit'' coupling to the total energy $cP_{0}$ given by (\ref{6.6p}). Thus, noting (\ref{6.6v}) as well, we have
\begin{equation} 
q_0=\dot{S}\bdot (v\wedge\gamma_{0})
=(i\dot{v}v)\bdot (s\wedge\gamma_{0})
= (\ddot{\r}\boldsymbol{\times} \dot{\r})\bdot\s.\label{6.47b}
\end{equation}
And for the vector part 
we get
\begin{align} 
\mathbf{q} &=(\dot{S}\bdot v)\wedge\gamma_{0}
=i(s\wedge\dot{v}\wedge v)\bdot \gamma_{0}
\notag \\
&=\dot{\gamma}\,\s\boldsymbol{\times} \dot{\r}
+\gamma\,\ddot{\r}\boldsymbol{\times} \s_{\perp}. \label{6.47c}
\end{align}
This concludes our synopsis of spacetime splits.

\subsection{Quantization of Stationary States}

In this section we examine the quantization of stationary states for a Pilot Particle, that is, a point particle with spin and clock. 
As defined by (\ref{6.26}) the \textit{particle orbit} $\r=\r(\tau)  $ is a projection of the particle path $ z=z(\tau) $ into an inertial system. The orbit is bounded and periodic if $\r(\tau)=\r(\tau+\tau_{n})  $ for some period $\tau_{n}$. For a  \textit{stationary state}, we also require periodicity of the particle's comoving frame:
\begin{equation}
e_{\mu}(\tau+\tau_{n})=e_{\mu}(\tau),
\label{6.48}
\end{equation}
and constant energy 
\begin{equation}
E=cP\bdot \gamma_{0}-m_{e}c^{2}.
\label{6.49}
\end{equation}
A somewhat stronger periodicity condition that implies (\ref{6.48}) is compactly expressed in terms of the rotor   
\begin{equation}
\hat{\Psi}(\tau_{n})=R(\tau_{n})e^{-\i\varphi(\tau_{n})}=\hat{\Psi}(0).\label{6.50}
\end{equation} This implies  periodicity of the wave function phase, which we now consider in detail.

Here we employ quantization conditions restricted to the particle path $z=z(\tau)$, so  we have the so-called \textit{Action integral}
\begin{equation}
\oint^{\tau_{n}} P\bdot dz=\oint(p+\frac{e }{c}A)\bdot dz=0
\label{6.51}
\end{equation}
where, from (\ref{6.6r}), $p=m_{e}c v+\dot{S}\bdot v$ is the particle momentum and $A$ is the external vector potential.

As before, the split $ P\gamma_{0}=P_{0}+\mathbf{P} $  gives us
\begin{equation}
 \oint^{\tau_{n}} P\bdot v\, d\tau= \int_{0}^{T_{n}}P_{0}\,dt -\oint \mathbf{P}\bdot d\r=0, \label{6.51a}
\end{equation}
The constant energy condition for a stationary state reduces this to
\begin{equation}
P_{0}T_{n}=\oint \mathbf{P}\bdot d\r\quad \hbox{where}\quad
T_{n}=\oint \gamma d\tau.\label{6.52}
\end{equation}
And the quantized phase  of the wave function is expressed as the spatial quantization condition
\begin{equation}
\oint \mathbf{P}\bdot d\r
=\oint (\p+\frac{e }{c}\A)\bdot d\r=(n+\half)h,
\label{6.53}
\end{equation}
where $ n $ is a positive integer.
It follows that temporal quantization is given by
\begin{equation}
P_{0}T_{n}=\int_{0}^{T_{n}} (p_{0}+\frac{e }{c}A_{0}) dt= (n+\half)h.\label{6.54}
\end{equation}
The originator of this equation, E. J. Post \cite{Post82}, called it  \textit{Electroflux quantization}. However, we see it as setting the period of the electron clock, which is then coordinated with quantization of orbit integrals
in (\ref{6.51a}).

Equation (\ref{6.53}) is well known in standard quantum mechanics as a \textit{WKB approximation}, but here it is exact, and, as we shall see,  sufficient for exact results from the PPM.
There has been some dispute about significance of the term $\half h$, which is sometimes identified as a \textit{zero point energy}. We have identified it more specifically as a consequence of electron zitter, so we can consistently ignore it when we ignore zitter. But we should not forget its likely role in the Hall effect. 

Now note that
\begin{equation}
q\bdot dz=(\dot{S}\bdot v)\bdot dz=(\dot{S}\bdot v)\bdot v\,d\tau=0.\label{6.51b}
\end{equation}
Hence, the spin momentum $q$ does not contribute to the action integral on the complete particle path. However, the split
\begin{equation}
q\bdot v =<q \gamma_0 \gamma_0 v> =\gamma q_0 -\mathbf{q}\bdot \dot{\r}=0.\label{6.51c}
\end{equation}
shows that there could be equal and opposite contributions to the spacelike and timelike integrals (\ref{6.53}) and (\ref{6.54}).
Their integrals need not cancel, however. Indeed,
we shall see good reason to assume
\begin{equation}
\oint \mathbf{q}\bdot d\r=0
\label{6.51d}
\end{equation}
while $q_0$ survives in the timelike integral.

Further conditions on the action integrals reflect symmetries and conservation laws. Central symmetry in an atom can be expressed by locating the origin of the electron position vector $\r=\z-\x_0$ at the nucleus.
Then conservation of angular momentum $\mathbf{L}=\r\wedge\p=i(\r\boldsymbol{\times}\p)$ is expressed by separating radial and angular motions. Accordingly, we introduce the split
\begin{align}
\p \bdot d\r&= <\p \r \r^{-1}d\r >
=(\p \wedge \r)\bdot(\hat{\r}\wedge d\hat{\r}) 
+\p\bdot \hat{\r}\,dr\notag\\
&=-\mathbf{L}\bdot(\i\,d\phi)+\p\bdot \hat{\r}\,dr
=|\mathbf{L}| d\theta +p_r dr,
\label{6.51e}
\end{align}
where $dr=\hat{\r}\bdot d\r$ and $\hat{\r}\wedge d\r=\i d\,\theta$. Then we have
\begin{equation}
\oint \p \bdot d\r
=\oint |\mathbf{L}| d\theta +\oint p_r dr=nh,
\label{6.51f}
\end{equation}
with
\begin{equation}
\oint |\mathbf{L}| d\theta =2\pi |\mathbf{L}| = {\ell} h,
\label{6.51eg}
\end{equation}
where the \textit{angular momentum quantum number} ${\ell}<n$ is a non-negative integer.
It follows that the radial motion must also be quantized with
\begin{equation}
 \oint p_r dr=2\pi  p_r=n_r h,\label{6.51eh}
\end{equation}
where $n_r=n-{\ell}$ is the \textit{radial quantum number}.

Symmetry with respect to a fixed direction in space, say $\bsig_3$ introduces a further quantum condition called ``space quantization:'' 
\begin{equation}
\oint(\r\boldsymbol{\times}\p) \bdot \bsig_3(\bsig_3 \bdot d\r) = m_{\ell} h  ,
\label{6.51fa}
\end{equation}
where $m_{\ell} \leq {\ell}$ is also a non-negative integer.

Now we consider important examples of PPM quantization.

\subsubsection{Relativistic Landau Levels}

For an electron in a constant magnetic field $ F=i\B=i\bsig_{3}B $, the spin $ \s=\pm\frac{\hbar }{2}\bsig_{3} $ is a constant collinear with $ \B $, and the orbit is determined by the equation of motion 

\begin{equation}
c\dot{\p}= \frac{e}{c}\dot{\r}\boldsymbol{\times} \B=\p\boldsymbol{\times} \bm{\omega}_{c},
\label{6.55}
\end{equation}
where $\omega_{c} =(e/m_{e}c)B $ is the cyclotron frequency.
This integrates immediately to 
\begin{equation}
\p= \p_{\perp} + \p_{\parallel},\label{6.56}
\end{equation}
where
\begin{equation}
\p_{\perp}= \frac{e}{c}\,\r\boldsymbol{\times} \B  =m_{e}c\,\r\boldsymbol{\times} \bm{\omega}_{c}\label{6.57}
\end{equation}
and the integration constant $ \p_{\parallel} $ is parallel to the magnetic field.
The vector potential for constant $ \B $ supplies a momentum 
\begin{equation}
 \frac{e}{c}\A=\frac{e}{2c}\B\boldsymbol{\times} \r  =\frac{m_{e}c}{2}
 \bm{\omega}_{c}\boldsymbol{\times}\r =-\frac{1}{2}\p_{\perp}\label{6.58}
\end{equation}
Hence, inserting the canonical momentum
\begin{equation}
\mathbf{P} =\p+\frac{e }{c}\A
=\frac{1 }{2} \p_{\perp}+\p_{\parallel}
\label{6.59}
\end{equation}
into the quantization condition (\ref{6.53}) gives us
\begin{equation}
\oint \mathbf{P}\bdot d\r
=\frac{1 }{2}\oint \p_{\perp} \bdot d\r=nh,
\label{6.60}
\end{equation}
Using (\ref{6.57}), we evaluate the integral
\begin{equation}
\frac{1 }{m_{e}c}\oint \p_{\perp} \bdot d\r=\oint \bm{\omega}_{c}\bdot (d\r\boldsymbol{\times} \r)=2\pi r^{2}\omega_{c}.
\label{6.61}
\end{equation}
Hence 
\begin{equation}
\p_{\perp}^{2}= (m_{e}c)^{2}\omega^{2}_{c}r^{2}
=2m_{e}c\hbar\omega_{c}n.
\label{6.62}
\end{equation}
Thus we obtain the quantized energy states for an electron in a magnetic field:
\begin{equation}
P_{0}
=\pm\left[m_{e}^{2}c^{2}+\p_{\parallel}^{2}+2m_{e}c\hbar\omega_{c}n\right]
^{\half}
\label{6.63}
\end{equation}
for $( n=1, 2, . . .) $. These are the \textit{relativistic Landau levels} first found by Rabi \cite{Rabi28} by solving the Dirac equation.

\subsubsection{Hydrogen Atom}

The hydrogen atom and its spectroscopy played a pivotal role in the development of quantum mechanics, including the transition from ``\textit{Old Quantum Theory}" (OQT), based on  quantization of particle orbits, to ``Wave Mechanics," which many claim denies the existence of orbits. The high point of OQT was Sommerfeld's relativistic formula for the  energy levels $ E_{n\ell} $ of hydrogen \cite{Ruark30}: 
\begin{equation}
\frac{E_{n\ell}}{m_{e}c^{2}} +1=\left[1+\frac{\alpha_{e}^{2}}{\left[n_{r}+\sqrt{n_{\ell}^{2}-\alpha_{e}^{2}}\:\right]^{2}} \right]  ^{-\half} ,\label{6.64}
\end{equation}
where the \textit{principle quantum number}
$ n=1, 2, . . . $ is the sum $ n=n_{r}+n_{\ell} $ of a \textit{radial quantum number} $ n_{r} $ and an \textit{angular quantum number} $ n_{\ell} $, while  $ \alpha_{e}$ is the fine structure constant.

On the other hand, successful treatment of hydrogen with the Dirac equation was a great triumph of Wave Mechanics! 
Considering the apparent deep philosophical divide between old and new quantum theories, it has been a perennial puzzle that the hydrogen energy spectrum derived from the Dirac equation \cite{Greiner90a} is identical in form to Sommerfeld's formula, with only a small difference in specification of the angular quantum number. The chief discrepancy has been resolved recently by Bucher 
\cite{Bucher08,Bucher06} who restored solutions with zero angular momentum that Sommerfeld had dismissed on grounds that they would collide with the atomic nucleus. Bucher called these solutions ``Coulomb oscillator" states and demonstrated that they have potential for deepening our understanding of chemical bonds \cite{Bucher07}. The problem remained to account for spin, as that had not been considered in OQT. The pilot particle model (PPM), regarded as an extension of OQT to include spin, is offered here as the solution.
Since quantization rules for the PPM are derived from the Dirac equation, we should be able to resolve any discrepancies between Old and New Quantum Theory. Let's see how.

The theoretical apotheosis of OQT
was Max Born's book: \textit{The Mechanics of the Atom} \cite{Born27}. Ironically, it was hardly off the press when Born collaborated with Heisenberg and Jordan to eclipse OQT completely  with the new ``Matrix Mechanics.'' Even so, Born boldly projected a second edition that would bring the theory to a satisfactory conclusion. That never happened, although he published quite a different kind of book with Jordan in its place. Even so, the present work can be regarded as revitalizing the original OQT program by linking it to the ascendant Dirac equation. 
The OQT method for solving the Hydrogen atom in Born's book applies equally well here, so details leading to the Sommerfeld formula need not be repeated (See also \cite{Ruark30}). A new approach to solving the equations of motion is presented in the next subsection.  Here we concentrate on issues of interpretation,
especially concerning the introduction of spin in the conservation laws for momentum 
\begin{equation}
p=m_{e}cv+ \dot{S}\bdot v=P-\frac{e }{c}A ,\label{6.65}
\end{equation}
and angular momentum. 

Applying (\ref{6.6p}) and (\ref{6.47a}),
a spacetime split of the momentum gives us an expression for the total energy
\begin{equation}
 cP_{0}=E+ m_{e}c^{2}=m_{e}c^{2}\gamma-V
 +c(\ddot{\r}\boldsymbol{\times} \dot{\r})\bdot\s, \label{6.66}
\end{equation}
where
$  V=eA\bdot\gamma_{0}$ is the Coulomb potential
and the last term looks like a spin-orbit energy. 
Spin-orbit coupling is treated as a perturbation even in standard Dirac theory, so let us omit it for the time being.
Without that explicit spin contribution, the energy form used by Sommerfeld is identical to the one used in Dirac theory. Our problem therefore is to resolve any discrepancies in interpretation.

For constant energy, Sommerfeld solved the equations of motion for the particle orbit, which turns out to be a precessing ellipse. Then he made the orbit periodic by imposing the quantization condition (\ref{6.53}), and  decomposed it into angular and radial quantum  conditions using spherical coordinates $(r,\theta,\varphi)$ with coordinate frame $(\hat{\r},\, \e_{\theta},\,\e_{\varphi})$. Thus,
\begin{equation}
\oint \mathbf{p}\bdot d\r = nh
=(n_{r}+n_{\theta} +n_{\varphi})h,
\label{6.68}
\end{equation}
with radial quantization 
\begin{equation}
\oint \mathbf{p}\bdot \hat{\r}\,dr= \oint p_{r}dr= n_{r}h,
\label{6.69}
\end{equation}
angular momentum  quantization 
\begin{equation}
 \oint \mathbf{p}\bdot
(\e_{\theta}\,d\theta+\e_{\varphi\,}d\varphi)
=  n_{\theta}h+n_{\varphi}h
\label{6.70}
\end{equation}
with $n_{\theta}+n_{\varphi}=n_{\ell}$,
and space quantization 
\begin{equation}
 \oint \mathbf{p}\bdot
\e_{\varphi\,}d\varphi=\oint p_{\varphi}d\varphi
= n_{\varphi}h
\label{6.70a}
\end{equation}
with $n_{\varphi}=m_\ell $, in agreement with the conditions following (\ref{6.53}).

Following Bucher \cite{Bucher08}, we replace Sommerfeld's assumption that $ n_{r}=0 $ by $ n_{r}=1 $ in the ground state, and note from (\ref{6.68}) that the range of angular quantum numbers $ n_{\ell}=\ell $ must be changed to $ \ell=0, 1, 2,. . .,n-1 $. This change is indicated by referring to $\ell$ as the \textit{orbital} angular momentum quantum number. As a consequence,  circular orbits (for which $n_{\ell}=n)$ are excluded, but the ground state is a linear oscillator passing through the nucleus.

Now, in agreement with Dirac theory, for a pilot particle we have
\begin{equation}
n_{\ell}=(j+\half), \label{6.71}
\end{equation}
where $ j=\ell\pm\half $ is the \textit{total} angular momentum quantum number. Pilot theory gives this a clear meaning:
As the particle traverses a path the ``clock hand" $ e_{1} $ rotates in the spin plane, and the integer $ n $ is the number of complete cycles it makes in an orbital period, while the integer $ {n_{r}} $ is the number of cycles in a complete radial oscillation. But, for a given orbit, there are two distinct angular cycles for the clock hand: ``spin up," where $ j=\ell+\half $ so $ n_{\ell} = \ell+1 $, where orbital motion adds a half cycle to the clock hand; and ``spin down," where $ j=\ell-\half $ so $ n_{\ell} = \ell $, where orbital motion subtracts a half cycle from the clock hand. In each case there is an integral number of clock cycles in an orbital period.

This completes our reinterpretation of the Sommerfeld formula. But there is more, as we have not yet taken angular momentum conservation into account.
According to (\ref{6.44}), for an electron moving with transverse spin in a Coulomb field the total angular momentum $ \mathbf{j}=\mathbf{l}+\s $ is conserved, 
so we take that as a fixed axis defined by $\mathbf{j}=|\mathbf{j}|\bsig_{3} $. 
It follows from the quantization conditions (\ref{6.70}) and (\ref{6.70a})     
that $ \mathbf{l} $ and $ \s $ must precess around $\mathbf{j}  $ with 
a constant angular frequency $ \bm{\omega}_{p} =\omega_{p}\bsig_{3}$, and
inclined at a fixed angle given by 
\begin{equation}
\mathbf{l}\bdot\mathbf{j}=(j^{2}+{\ell}^{2}-(1/2)^{2}) \hbar^{2}/2 =j m_{\ell}\hbar^{2}, \label{6.75}
\end{equation}
where $ m_{\ell}= 0, \pm 1,  \pm 2, ..., \pm (\ell-1),  \pm \ell $. Though $ m_{\ell}$ is commonly known as the magnetic quantum number because it describes splitting of spectral lines in a magnetic field, it shows up spectroscopically even in the absence of a detectable magnetic field. Here we see  it arising from quantization of free precession of orbital angular momentum balanced by spin precession. Though its effect is not implicit in the Sommerfeld formula (\ref{6.64}), it does appear in spectra, which are statistical averages of light emitted from an ensemble of atoms in different orientations.
 
This insight helps us reconcile our particle model with the strange expectation values for angular momentum operators in wave mechanics, such as
\begin{equation}
\left\langle L \right\rangle=0,\quad  \text {but}\quad\left\langle L^{2} \right\rangle^{\half}=\sqrt{\ell(\ell+1)}\hbar^{2}   .\label{6.77}
\end{equation}
As number of authors \cite{Feynman65,McGervy81,Post95} have noted, they can be derived from expectation values of the standard operators $ L_{z} $ with eigenvalues $ \ell\hbar $ taken to be outcomes of measurement.
While others present this fact as a mere curiosity, Post \cite{Post95} has argued that the derivation can be regarded as a statistical average over an ensemble of equally probable measurement outcomes. 

Thus, for a given atom, the angular momentum aligned along the $ z-$direction can have any one of the $ 2\ell+1 $ values: $ +\ell\hbar, +(\ell-1)\hbar, ...,-(\ell-1)\hbar,-\ell\hbar.$ Regarding all values as equally probable, the average value for the squared angular momentum is therefore
\begin{align}
\left\langle L_{z}^{2} \right\rangle= \frac{\hbar^{2}}{2\ell+1}\sum_{n=-\ell}^{\ell}n^{2}&=
\frac{\ell(\ell+1)(2\ell+1)/6}{2\ell+1}\notag\\
&=\ell(\ell+1)\hbar^{2}/3 \label{6.78}
\end{align}
Since no direction is preferred, we have
\begin{equation}
\left\langle \mathbf{L}^{2} \right\rangle=
\left\langle L^{2} \right\rangle=\left\langle L_{x}^{2} \right\rangle+\left\langle L_{y}^{2} \right\rangle+\left\langle L_{z}^{2} \right\rangle=\ell(\ell+1)\hbar^{2},\label{6.79}
\end{equation}
where $\mathbf{L}=\r\wedge\p$ is as defined in the PPM.
For the total angular momentum with half integral quantum numbers the same argument gives a result with the same form.

This argument supports the view of Post \cite{Post95} and others that standard quantum mechanics is about statistical ensembles rather than individual physical systems.
For a more general argument, we need to explain how the PPM relates to the probability interpretation of the Dirac wave function.
A definitive relation must exist, because both are grounded in the Dirac equation. It should be remembered that the probability interpretation, for which Born received the Nobel prize in 1954 \cite{Born54,Pais82}, solved the problem of explaining spectral intensities, which OQT had failed to do. So we cannot dispense with probabilities, but we need to relate them to ensembles of particle states.
How to do that, is left here as an open problem, but with  suggestions for attacking it below. It should be noted that the quantization conditions  we ``derived'' from an integrability condition have been identified with those of OQT derived from Hamilton-Jacobi theory. This purported equivalence should be turned into a theorem. It suggests that H-J theory can be applied to rigorously define the approach to wave function singularities, thereby deriving particle equations of motion for them. It seems that a good start has already been made by Synge \cite{Synge54}, who used relativistic H-J theory in a unique way to solve the Hydrogen atom without Dirac theory. On reformulating that approach with STA, it should be evident how to incorporate spin and relate it to the PPM. This could be a helpful step toward relating the PPM to statistical ensembles.  
Related remarks will be made at the end of this paper.

We still have issues to address concerning the role of spin momentum in the hydrogen atom solutions.
Since the spin momentum $q$ was derived from Dirac theory, it is presumably embedded in solutions of the Hydrogen atom, though that is not at all evident in the standard Darwin solutions \cite{Greiner90} of the Dirac equation or the Sommerfeld energy spectrum (\ref{6.64}).
To study its role, we begin with the explicit appearance of the spin-orbit term
$q_0=c(\ddot{\r}\boldsymbol{\times} \dot{\r})\bdot\s$ of the total energy (\ref{6.66}).
To solve for the energy, we need to express this term as a function of position using the equation of motion 
\begin{equation} 
\dot{\p}=m_e c\ddot{\r}+\dot{\mathbf{q}}
= -\gamma\boldsymbol{\nabla} V
=-\gamma\r\frac{\partial_rV}{r}\label{6.80}
\end{equation}
For $\mathbf{q}=0$ we get a well known expression for the spin-orbit energy:
\begin{equation}
(\ddot{\r}\boldsymbol{\times} \dot{\r})\bdot\s
=\gamma \s\bdot(\r\boldsymbol{\times} \dot{\r})
\frac{\partial_rV}{m_ecr}
=  \mathbf{l}\bdot \s_{\perp}
\frac{\partial_rV}{m_e^2c^2r}.
 \label{6.81}
\end{equation}
Including $\mathbf{q} \neq 0$ adds unfamiliar terms without precedent or evident physical meaning.
But (\ref{6.51c}) tells us that we cannot dispense with  $\mathbf{q} $ entirely without killing the spin-orbit energy $q_0$, so we are put in a quandary. 

One way out of the quandary might be to assume that $\mathbf{q}$ does not contribute to quantizing  orbital motion by asserting the integral condition (\ref{6.51d}). That leads us to ignore $q_0$ to see what role $\mathbf{q}$ might play in Sommerfeld's original formulation: 
\begin{equation}
 cP_{0}=E+ m_{e}c^{2}=m_{e}c^{2}\gamma-V
  \label{6.82}
\end{equation}
We look for it in the $\gamma=c\dot{t}$ term. 
That term is related to spatial momentum by assuming 
\begin{equation}
v^2=\gamma^2-\dot{\r}^2 =1\quad\hbox{and}\quad\p=m_ec\,\dot{\r}
  \label{6.83}
\end{equation}
to get
\begin{equation}
m_ec\,\gamma=\sqrt{m_e^2c^2+\p^2}. 
  \label{6.84}
\end{equation}
Then quantization is imposed by inserting
\begin{equation}
\p^{2}=p_{r}^{2}+\frac{\mathbf{l}^2}{r^2}
=\left(n_{r}^{2}+\frac{\ell^2}{r^2}\right)\hbar^2  \label{6.85}
\end{equation}
With these assumptions, the energy (\ref{6.82})
becomes an explicit function of the radius $r$ which Sommerfeld solved to get an equation for the orbit. 

The point to be emphasized here is: though we know that the quantized momentum in (\ref{6.85}) could  include spin momentum, that possibility is excluded tacitly in (\ref{6.84}) and explicitly in (\ref{6.83}). 
So, to help resolve our spin-momentum quandary, we turn to a more geometrically perspicuous solution of Sommerfeld's problem that shows how spin can be incorporated to implement space quantization.

\subsubsection{The Kepler-Sommerfeld problem}

Sommerfeld solved the relativistic Kepler problem for Hydrogen to find closed electron orbits with quantized energies.  
Let's call this the \textit{Kepler-Sommerfeld} (K-S) problem. With the rise of standard Quantum Mechanics it was dismissed as a historical artifact. However,
the insight that Sommerfeld orbits are inherent in singular solutions of the Dirac equation revitalizes the K-S problem and restores it to a central place in relativistic particle physics.
 
Here we reconsider the K-S problem with a new spinor method enabled by Geometric Algebra. 
We reconstruct Sommerfeld's solution with the aim of generalizing it to include electron spin and  a foundation for perturbation theory.

To emphasize geometric structure, we adopt $c=1$ (in this Section only) and write  the equation of motion (\ref{6.80}) in the form
\begin{equation} 
\ddot{\r}=-\dot{t}\,\hat{\r}\,\partial_rV
+\mathbf{f},\label{6.90}
\end{equation}
where $V=k/r$, with $k=e^2/m_ec^2$, is the Coulomb potential. 
A (scaled) perturbing force $\mathbf{f}$ has been included for generality, but we shall ignore it until final remarks.

A first integral for the equation of motion can be determined from constants of motion. Accordingly, we rescale the \textit{energy constant} 
(\ref{6.82}) to
\begin{equation}
W\equiv cP_{0}/ m_{e}c^{2}=E/ m_{e}c^{2}+1=\dot{t}-V.
  \label{6.91}
\end{equation}
Another constant of motion is the \textit{angular momentum} (per unit mass) $\mathbf{L}=i\mathbf{l}$, defined by
\begin{equation} 
\mathbf{L}=\r\wedge\ddot{\r}=r^2\,\hat{\r}\,\dot{\hat{\r}}.
\label{6.92}
\end{equation}
This can be solved for
\begin{equation}
\dot{\hat{\r}}=\frac{\hat{\r}\,\mathbf{L}}{r^2}
\qquad \hbox{and} \qquad
\dot{\r}=\left(\dot{r}+\frac{\mathbf{L}}{r}\right)\hat{\r}. 
\label{6.93}
\end{equation}
Hence,
\begin{equation}
\dot{\r}^2=\dot{r}^2+\frac{l^2}{r^2},
\label{6.93a}
\end{equation}
where $l =|\mathbf{l}|=|\mathbf{L}|$.
Finally, on multiplying (\ref{6.90}) by $\mathbf{L}$ and using (\ref{6.93}) we get a first integral in the form
\begin{equation} 
\mathbf{L}\,\dot{\r}
=k(\hat{\r}+\mathbf{e}).\label{6.94}
\end{equation}
This is an equation  relating velocity and focus radius vector for a precessing ellipse with \textit{eccentricity vector} $\mathbf{e}$ of constant magnitude.
Rather than integrating it for the orbit, we turn now to a different method.

\subsubsection{Spinor Particle Mechanics}

We follow the method in \cite{Hest98} to represent the electron's particle orbit $\r=\r(\tau)$ by a spinor $U=U(\tau)$ defined by
\begin{equation}
\r=U\bsig_1\tU =r \hat{\r}
\label{6.95}
\end{equation}
and introduce a new parameter $s=s(\tau)$ defined by
\begin{equation}
r=|U|^2 =\frac{d\tau}{ds}\qquad \hbox{so}\qquad \dot{s}=1/r.
\label{6.96}
\end{equation}
The proper time derivative of the ``position spinor'' $U$ can be parmeterized in general form
\begin{equation}
\frac{dU}{ds}=\half (\dot{r}r^{-1}-i\bm{\omega}_r)U.
\label{6.97}
\end{equation}
Then we can choose the angular velocity $\bm{\omega}_r$ so that $\r\bdot\bm{\omega}_r=0$ and
\begin{equation}
2\dot{U}\bsig_1\tU=
 (\dot{r}r^{-1}-i\bm{\omega}_r)\r=\dot{\r}
\label{6.98}
\end{equation}
Thus we find a simple relation between the spinor derivative and the particle velocity:
\begin{equation}
\frac{dU}{ds}=r\dot{U}=\half\dot{\r}\,U\bsig_1.
\label{6.99}
\end{equation}
Differentiating once more, we obtain
\begin{equation}
\frac{d^2U}{ds^2}=\half(\ddot{\r}\r+\half\dot{\r}^2)U.
\label{6.100}
\end{equation}
This completes our formulation of spinor kinematics by relating it to vector kinematics. 

To transform our equations of motion from vector variables to spinors, we use (\ref{6.96}) to put the energy constant of motion (\ref{6.91}) in the form
\begin{equation}
W=\dot{t}-k\dot{s}.
  \label{6.101}
\end{equation}
This integrates immediately to $W\tau=t+ks$.  Instead, we use it along with $\dot{t}^2-\dot{\r}^2=1$ to eliminate the $\dot{t}$ variable from the equation of motion (\ref{6.90}). Thus, with a little algebra to put it in a form for comparison with (\ref{6.100}), we obtain
\begin{equation} 
\ddot{\r}\r+\half(\dot{\r}^2-V^2)=\half(W^2-1).\label{6.102}
\end{equation}
Using (\ref{6.93a}), we see that 
\begin{equation}
\dot{\r}^2-V^2=\dot{r}^2+\frac{l^2-k^2}{r^2},
\label{6.103}
\end{equation} 
which tells us that the $V^2$ term can be absorbed into the angular momentum by shifting to a precessing system.
Accordingly, we factor the position spinor into
\begin{equation}
U=U_pU_r, \quad \hbox{where}\quad
U_p=e^{-\half i\bm{\omega}_p\tau}
\label{6.104}
\end{equation}
generates precession 
with a fixed angular velocity
\begin{equation}
\bm{\omega}_p=\frac{l^2-k^2}{l^2}\bsig_3.
\label{6.105}
\end{equation}
And substituting (\ref{6.102}) into (\ref{6.100})
gives us
\begin{equation}
\frac{d^2U_r}{ds^2}=\frac{1}{4}(W^2-1)U_r.
\label{6.106}
\end{equation}
This is the desired spinor equation of motion for the orbit. It describes an ellipse expressed as a 2-d harmonic oscillator. Thus, the complete position spinor $U=U_pU_r$ describes a precessing ellipse.

The solution of (\ref{6.106}) has the familiar general form:
\begin{equation}
U_r=\alpha\cos{\half\varphi}+\i\,\beta \sin{\half\varphi},\label{6.107}
\end{equation}
where $\varphi=\omega s$ with angular frequency $\omega=(W^2-1)^\half$.
It can be substituted into (\ref{6.95}) to get an explicit harmonic equation for the radius vector $\r=\r(s)$, but it is simpler to use as is.
At least that substitution relates the coefficients to the ellipse's semi-major, minor axes $(a,b)$, semilatus rectum $\Lambda$, and eccentricity $|\e|$ :
\begin{equation}
a=\alpha^2,\quad  b=\alpha\beta,\quad
\Lambda=\beta^2,\quad
\e^2=\frac{\alpha^2+\beta^2}{\alpha^2}.
\label{6.108}
\end{equation}
This completes our spinor solution of the relativistic Kepler problem.

Now  the solution can be quantized as before by applying the Sommerfeld quantization conditions. The resulting orbit parameters (labeled by the quantum numbers) are
\begin{equation}
a_{n}=n^2r_B,\quad  b_{n\ell}=n\ell r_B,\quad
\Lambda_{\ell}=\ell^2r_B,
\label{6.109}
\end{equation}
where $r_B=\hbar^2/m_ec^2$ is the Bohr radius.
The orbital frequency is given by 
\begin{equation}
\omega_{n\ell}=\sqrt{W_{n\ell}^2-1},\label{6.110}
\end{equation}
with quantized values determined by the energy $E_{n\ell}$ in Sommerfeld's formula (\ref{6.64}).

It must be understood that these quantized values cannot be directly compared to observed values in atomic spectra.
Instead, one must compare the averages over an ensemble of atoms with different spatial orientations derived in (\ref{6.78}) and (\ref{6.79}).
Bucher \cite{Bucher16} has pointed out that an ellipse's geometric properties of size and shape have independent quantitative representations by its major axis and semilatus rectum respectively, or, for a particle orbit, by total energy and angular momentum. Consequently, the major axis can be held constant while the angular momentum in (\ref{6.109}) is averaged to get   
\begin{align}
a_{n}=n^2r_B,\quad  &b_{n\ell}=n\sqrt{\ell(\ell+1)} r_B,\notag \\
&\Lambda_{\ell}=\ell(\ell+1)r_B,
\label{6.109b}
\end{align}
in agreement with Bucher \cite{Bucher08}.
Corresponding averages can by obtained by simple substitution in Eqs. (\ref{6.85}), (\ref{6.93a}), (\ref{6.103}) and (\ref{6.105}).

We can easily generalize the solution to include Sommerfeld's space quantization as a freely precessing ellipse with obital angular momentum balanced by spin in accordance with (\ref{6.75}).
The elliptical orbit is tilted and precesses with quantized projected area exactly as described by Bucher \cite{Bucher08}.
For a tilt angle with cosine $m_{\ell}=\hat{\mathbf{l}}\bdot\bsig_3$, the tilt of the angular momentum is given by
\begin{equation}
\hat{\mathbf{l}}=U_{m_{\ell}}
\,\bsig_3\,\tU_{m_{\ell}}=U^2_{m_{\ell}}\,\bsig_3.
\label{6.111}
\end{equation}
Hence the tilt rotor is
\begin{equation}
U_{m_{\ell}}=(\hat{\mathbf{l}}\,\bsig_3)^\half.
\label{6.112}
\end{equation}
And the position spinor (\ref{6.104}) generalizes to
\begin{equation}
U=U_pU_{m_{\ell}}U_r.
\label{6.113}
\end{equation}
This appears to be a satisfactory solution to the space quantization problem.
However, our quandary about including spin momentum to derive it remains unresolved --- at least for the time being.

Finally, incorporating the perturbing force $\mathbf{f}$ in (\ref{6.90})
into our spinor equation, we have
\begin{equation}
2\frac{d^2U}{ds^2}+i\bm{\omega}_pU-\half(W^2-1)U
=\mathbf{f}\r U=|U|^2\mathbf{f}U\bsig_1 .
\label{6.106a}
\end{equation}
Our spinor equation of motion for the relativistic Kepler problem is almost identical to the one for the non-relativistic Kepler problem, where it is known as the \textit{Kustanheimo-Stiefel} equation \cite{Hest98,Hest86}. The latter has been applied extensively to perturbations in celestial mechanics by Vrbik \cite{Vrbik10} with great effect.
We should expect it to be equally effective for treating perturbations in relativistic quantum mechanics. It might be further enhanced by spinor formulations of perturbations such as given in \cite{Hest74a}.

\section{Electron Zitter } \label{sec:}

Dirac's strong endorsement \cite{Dirac58} of Schr\"odinger's \emph{zitterbewegung} \cite{Schr30} as a fundamental property of the electron has remained unchallenged to this day, though it plays little more than a metaphorical role in standard quantum mechanics and QED.
However, evidence is mounting that zitterbewegung is a real physical effect, observable, for example, in Bose-Einstein condensates \cite{LeBlanc13} and semiconductors \cite{Zawadzki11}.  
Analysis with a variant of the model proposed here  \cite{Hest10} even suggests that zitterbewegung
has been observed already as a resonance in electron channeling \cite{Gouanere08}. That experiment should be repeated at higher resolution to confirm the result and identify possible fine structure in the resonance \cite{Gouanere15}.

Theoretical analysis of \emph{zitterbewegung}, or just \emph{zitter}, requires a formulation in terms of local observables.
We have already noted that the zitter frequency is inherent in the phase of the Dirac wave function.
But Schr\"odinger claimed more, namely, that it is to be interpreted as a frequency  of position oscillations at the speed of light about a mean velocity, and it has been further claimed that association of electron spin with circular zitter  was implicit in his analysis \cite{Huang52}. 

In Section IIIC we proved the existence of plane wave solutions of the Dirac equation that are fibrated by lightlike helical paths fixed at the zitter frequency $\omega_e$. This is a limiting case of conventional plane wave solutions of the Dirac equation, so it does not arise in in the construction of wave packets and other solutions of the Dirac equation, or the flow of observables in Section IIID.

Fortunately, the adjustment required to incorporate zitter into standard Dirac theory is fairly straightforward, so we can brief.
We suppose that the Pilot Particle Model (PPM) studied in the previous Section can be derived by averaging out high frequency zitter fluctuations. 
Accordingly, we define the \emph{``Zitter Particle Model'' }(ZPM) to restore those fluctuations.

More generally, we see that lightlike zitter velocity factors the Dirac Lagrangian into separate electron and positron parts. This has implications for QED, but that is outside the scope of this paper.

\subsection{Zitter Particle Model}

It should be recognized that the Dirac equation by itself does not imply any relation of the wave function to electron velocity.
Hence, a fundamental question in Dirac Theory is how to relate  observables in the Dirac equation to particle position or path.
That is sometimes cast as the problem of defining a position operator. The PPM  offers an implicit answer by modeling the electron as a particle with a timelike velocity $v=e_{0}=\dot{z}$. 
Instead, the ZPM in this Section models electron velocity as a lightlike vector and defines a complete set of local observables consistent with that. 

We have seen that the hand of the electron clock rotates with the zitter frequency, so it is natural to identify the velocity of circulation with the vector $e_{2}$ while $e_{1}$ is the direction of the zitter radius vector. Since there are two senses to the circulation corresponding to \textit{electron/positron}, we have two null vector  particle velocities:
\begin{equation}
e_{\pm}= v\pm e_{2}=R\gamma_{\pm}\tR,\quad\hbox{with}\quad\gamma_{\pm}
=\gamma_{0}\pm\gamma_{2}.\label{7.1}
\end{equation}
We restrict our attention to the electron case and redefine the local observables to incorporate zitter. Our choice of sign here is a convention in agreement with \cite{Hest10}.  

Accordingly, we define the electron's ``\emph{chiral velocity}" $u$ and ``\emph{chiral spin}"  $S$  by
\begin{equation}
u=R\gamma_{+}\tR=v+e_{2} \label{7.2}
\end{equation}
and
\begin{equation}
S=\frac{\hbar}{2}R\gamma_{+}\gamma_{1}\tR=ud,\quad \hbox{where}\quad d=\frac{\hbar}{2}e_{1}.\label{7.3}
\end{equation}
Note that the null velocity $u^{2}=0$ implies null spin bivector $S^{2}=0.$
Using the identities
\begin{equation}
ue_{1} =e_{0}e_{1}+ie_{0}e_{3}=iue_{3},\label{7.4}
\end{equation}
we can write $S$ in the several equivalent forms:
\begin{equation}
S=ud=v(d+is)=vd+\overline{S}=ius.\label{7.5}
\end{equation}
To designate the vector $d$, let me coin the term ``\emph{spinet}" (that which spins) as  counterpart of the ``spin" (vector) $s$, since they generate electric and magnetic moments together.  The overbar designates a ``\emph{zitter average}," that is,  an average over the zitter period $\tau_{e}=2\pi/\omega_{e}$. So the ``\emph{linear velocity}" $v=\bar{u}$ is an average of the chiral velocity $u$, and, since $\bar{d}=0$, the spin bivector $\overline{S}=isv$ is the zitter average of the \emph{chiral spin} $S$.

The spinor kinematics for the ZPM is a straightforward generalization of the PPM, and has been thoroughly studied in \cite{Hest10}, so the results are simply summarized here. 
As before, the rotational velocity for local observables is specified by 
\begin{equation}
\Omega =2\dot{R}\tR.\label{7.6}
\end{equation}
Generalizing (\ref{6.6a}) but ignoring the dummy parameter we assume the rotor equation of motion
\begin{equation}
\hbar\dot{R}\gamma_{+}\gamma_{1}= \frac{\hbar}{2}\Omega R\gamma_{+}\gamma_{1}= pR\gamma_{+}.\label{7.7}
\end{equation}
Hence,
\begin{equation}
\Omega S= pu.\label{7.8}
\end{equation}
As in the analogous PPM case, the bivector part of this expression gives us the spin equation of motion:
\begin{equation}
\dot{S}=\Omega\boldsymbol{\times} S= p\wedge u.\label{7.9}
\end{equation}
And the scalar part  gives us an expression for particle energy:  
\begin{equation}
p\bdot u=\Omega\bdot S > 0.\label{7.10}
\end{equation}
We cannot divide (\ref{7.8}) by the null vector $u$ to get an analogue of equation (\ref{6.6ff}), 
but we can divide by $v$ to get a comparable expression for momentum 
\begin{equation}
p=(p\bdot v)u+\dot{S}\bdot v .\label{7.10a}
\end{equation}
And (\ref{7.7}) also gives us 
\begin{equation}
\dot{u}=\Omega\bdot u = p\bdot S.\label{7.11}
\end{equation}
 
ZPM dynamics is driven by the obvious Lorentz force (\ref{6.6}) analogue:
\begin{equation}
\dot{p}=\frac{e}{c}F\bdot u,\label{7.12}
\end{equation}
which can be related directly to the rotor equation (\ref{7.7}) by derivation from a common Lagrangian, as shown in \cite{Hest10}.

The ZPM is not complete until we specify its kinematics relating  the particle velocity to its spacetime path.
Accordingly, we define  
\begin{equation}
r_{e}=\lambda_{e}e_{1}\qquad\hbox{with}\qquad 
\lambda_{e}=\frac{\hbar}{2m_{e}c}=c/\omega_{e}\label{7.13}
\end{equation}
as the radius vector for circular zitter at the speed of light.
The zitter center follows a timelike path $z=z(\tau)$ with velocity $v=\dot{z}$. Hence, the particle path $z_{+}=z_{+}(\tau)$ with lightlike velocity 
\begin{equation}
u=\dot{z}_{+}=v(\tau)+\dot{r}_{e}(\tau)\label{7.14}
\end{equation}
integrates to
\begin{equation}
z_{+}(\tau)=z(\tau-\tau_{c})+r_{e}(\tau),\label{7.15}
\end{equation}
where the time shift $\tau_{c}$ amounts to an integration constant to be determined below.
Note that the time variable is the proper time of the zitter center.

One satisfying feature of the ZPM is the physical interpretation it gives to $e_{1}$ and $e_{2}$, which are merely rotating vectors in the PPM. From (\ref{7.13})
see that $e_{1}=\hat{r}_{e}$ is the unit radius vector of the zitter, and from (\ref{7.2}) we see that $e_{2}=\dot{r}_{e}$
is the zitter velocity.

Remarkably, our model of the electron as a particle with circular zitter was proposed by Slater \cite{Slater26} well before the Dirac equation and Schr\"odinger's zitterbewegung.
His argument linking it to the null Poynting vector of the photon may also prove prophetic. Of course, we get much more than Slater could by embedding the model in Dirac theory.

\subsection{Zitter Field Theory}

We have seen that introduction of zitter into Dirac theory requires replacing the timelike velocity $v$ in the Dirac current with the lightlike velocity $u$. 
For decades I thought that, for consistency with Dirac field theory, this requires modification of Dirac's equation. 
I realized only recently that it is achieved 
more simply by a change in the wave function.
That suggests that radical surgery to excise the mass term from the Dirac equation, as required by the Standard Model, is unnecessary at best.
Here we introduce a new surgical procedure based on identification of electron and positron with zitter states of opposite chirality. 
The surgery is best done on the Lagrangian, because that is an invariant of the procedure. We shall see that it appears to vindicate Dirac's original idea of hole theory: the ``cut'' of the wave function to separate electron and positron is merely done in a different way.

Since the surgery must be gauge invariant, it is convenient to express the Lagrangian in terms of the gauge invariant derivative 
\begin{equation}
D\Psi=\gamma^{\mu}D_{\mu}\Psi = \gamma^{\mu}(\partial_{\mu}\Psi+\frac{e}{\hbar c}A_{\mu}\Psi i\gamma_{3}\gamma_{0}), \label{E.1}
\end{equation}
with due attention to the caveat about the gauge invariant derivative expressed in connection with (\ref{4.39}).

Whence the Dirac Lagrangian (\ref{B.5r}) takes the compact form
\begin{equation}
\mathcal{L}=\left\langle\hbar\,D\Psi i\gamma_3\widetilde{\Psi}
- m_{e}c\Psi\widetilde{\Psi}
\right\rangle.\label{E.1a}
\end{equation}
Now, note that the two null vector velocities introduced in (\ref{7.1}) correspond to distinct \emph{Majorana states}:
\begin{equation}
\Psi_{\pm}=\Psi(1\pm\bsig_2)=\Psi\gamma_{\pm}\gamma_{0},\label{E.2}
\end{equation}
which are related by charge conjugation:.
\begin{equation}
\Psi(x) \quad \rightarrow \quad \Psi_C(x)=\Psi(x)\bsig_1 \label{E.2a}
\end{equation}
	Accordingly, we identify $\Psi_{+}$ with the electron and $\Psi_{-}$ with the positron.

Now, a ``Majorana split" $\mathcal{L}=\frac{1}{2}(\mathcal{L}_{+}+\mathcal{L}_{-})$ of the Lagrangian (\ref{E.1a}) gives us separate Lagrangians for positron and electron:
\begin{multline}
\hspace*{.2in}\mathcal{L}_{\pm}=\left\langle[\hbar D \Psi_{\pm} i\gamma_3- m_{e}c\Psi_{\pm}]\tPsi\right\rangle\\
=\left\langle[\hbar D \Psi i\gamma_3 - m_{e}c\Psi]\gamma_{\pm}\gamma_{0}\tPsi\right\rangle.
\qquad\qquad\label{E.3}
\end{multline}
Variation with respect to $\tPsi$ gives us charge conjugate Dirac equations
\begin{equation}
D\Psi_{\pm} i\bsig_3- \frac{\omega_{e}}{2c}\Psi_{\pm}\gamma_{0}=0,\label{E.4}
\end{equation}
To emphasize the incorporation of chiral zitter, let's call them ``\emph{chiral Dirac} equations." The last term has been expressed in terms of zitter frequency instead of electron mass to emphasize that the two signs refer to the opposite chiralites of electron and positron.

Obviously, the chiral Dirac equations (\ref{E.4}) differ from the original Dirac equation  only by the simple projections (\ref{E.2}). There is a significant difference, though, in the appropriate choice of observables for each solution, so our analysis will be facilitated by carefully defining the key observables. We concentrate on the electron, as the positron analog is obvious.

Chiral projection produces a small but physically significant difference in the energymomentum tensor (\ref{C.6}), so it is worth spelling out.
As before, from translational invariance of the chiral Dirac Lagrangian (\ref{E.3}) we derive the chiral energymomentum tensor
\begin{equation}
\overline{T}_{+}(n)=\gamma^{\mu}
\hbar\left\langle D_{\mu}\Psi_{+}i\gamma_{3}\tpsi n\right\rangle
.\label{E.10}
\end{equation}
As before, the flux in the direction of $v$ is especially significant for two reasons: First, because it is related to the electron clock by
\begin{equation}
\overline{T}_{+}(v)=\gamma^{\mu}
\hbar\left\langle D_{\mu}\Psi_{+} i\bsig_{3}\widetilde\Psi \right\rangle=
-\frac{\hbar}{2}\gamma^{\mu}\rho(u\bdot D_{\mu}e_{1})
.\label{E.11}
\end{equation}
Second, because its values are determined by the chiral Dirac equation, with the result:
\begin{equation}
\overline{T}_{+}(v)=m_{e}c\rho u-D\bdot(\rho S),\label{E.12}
\end{equation}
where
\begin{equation}
\rho S=\hbar\Psi_{+}i\bsig_{3}\widetilde\Psi_{+}=\half\hbar\Psi\gamma_{3}\gamma_{+}\widetilde\Psi.  \label{E.12a}
\end{equation}
The zitter average of this tensor is obviously the tensor $\overline{T}(v)$ given by
(\ref{B.7r}), where the ``\emph{linear momentum}"  $m_{e}cv$ is the zitter mean of the ``\emph{chiral momentum}"  $m_{e}cu$.

For the energy density we get
\begin{equation}
v\bdot \overline{T}_{+}(v)=m_{e}c\rho-(v\wedge D)\bdot (\rho S).\label{E.13}
\end{equation}
The last term on the right can be written
\begin{equation}
(v\wedge D)\bdot (\rho S)=\D\bdot (\rho\d)=\boldsymbol{\nabla}\bdot(\rho \d)
-\frac{e \rho}{\hbar c}\A\bdot\d,\label{E.14}
\end{equation}
where $\d=dv$. The divergence does not contribute to the total energy by Gauss's theorem. Locally, of course, the whole term fluctuates with the zitter frequency as $\d$ rotates, so its zitter average vanishes.
This completes our characterization of local observables in chiral Dirac theory.

More details are needed to flesh out the picture.
However, the bottom line is, whether or not zitter arises from an orbital motion,  zitter is still manifested in the phase of the electron wave function and, thereby, in the quantization of electron states.

\section{Ontology cum Epistemology} \label{sec:X}


When long-standing scientific debates are finally resolved, it invariably turns out that both sides are correct in positive assertions about their own position but incorrect in negative assertions about the opposing position.

The Great Debate over the interpretation of quantum mechanics can be cast as a dialectic between Einstein's emphasis on ontology and Bohr's emphasis on epistemology \cite{Jaynes96}. 
This paper offers a new perspective on the debate by focusing on the local observables determined by the Dirac wave function.

As we saw in  Section IV, the spacetime flow of local observables is governed by the equation
\begin{equation}
\rho(P-\frac{e }{c}A)=m_{e}c\rho u-\square\bdot(\rho S),\label{8.20}
\end{equation}
which, with some provisos, can be regarded as a reformulation of the Dirac equation.  When restricted to a Dirac streamline,  as an electron path $z=z(\tau)$ with proper velocity $u=\dot{z}$, it becomes a relativistic generalization of the de Broglie-Bohm guidance equation.
Thus it provides a secure foundation for a Pilot Wave interpretation of the Dirac wave function, wherein the scalar $\rho$ is interpreted as a density of particle paths.

Since the Dirac equation is a linear differential equation, the superposition principle can be used to introduce probabilities in initial conditions and construct the wave packets of standard quantum mechanics with the Born interpretation of $\rho$ as probability density. This establishes full compatibility between the Pilot Wave and Born interpretations of Dirac wave functions. 
 
A definitive analysis of zitter in the Dirac wave function was given in Section IVC and proposed as a defining property of the electron. That shows that zitter can be regarded as electron phase incorporated in particle motion. It is included in the Pilot Wave  guidance law with a slight modification of (\ref{8.20}) given by (\ref{4.44}). That completes the description of the electron in Born-Dirac theory as a particle with intrinsic spin and zitter in its motion.

Concerning the Born rule  for interpreting the wave function in Quantum Mechanics:
There is no doubt that probability is essential for interpreting  experiments. 
Indeed, overcoming the failure of Old Quantum Theory to account for intensities of spectral lines was one of the first great victories for Quantum Mechanics and Born's rule for statistical interpretation. But that did not seal the demise of Old QT as commonly believed. For Old QT is resurrected in Section  V and revitalized as a particle model for the Hydrogen atom. Moreover, it comes not to destroy QM but to fulfill! For the particle model offers an ontic interpretation to QM, while QM offers an empirically significant way to assign probabilities to particle states. The particle model provides electron states with definite position, momentum, spin and (zitter) phase. 

The Born rule offers a way to assign probabilities to these states. However, such probabilities do not imply uncertainties inherent in Nature as often claimed. Rather, they express limitations in our knowledge and control of specific states, best described by  Bayesian probability theory so ably expounded by E. T. Jaynes \cite{Jaynes96,Jaynes03}.

Of course, the present approach calls for reconsideration of many arguments and applications of standard quantum mechanics, especially those involving zitter and the Heisenberg Uncertainty Relations, which are already burdened by many conflicting interpretations \cite{Bunge56}.
 
A new perspective on the Great Debate on the interpretation of QM will be introduced in a sequel to the present paper \cite{Hest19b}.

\vspace{0.2 cm}
\noindent \textbf{Acknowledgment.}  I am indebted to Richard Clawson and Steve Gull for critical give and take about Real Dirac Theory over many years.

\bibliographystyle{unsrt}
\bibliography{zbw}

\end{document}